\documentclass[prb,twocolumn,showpacs,floatfix]{revtex4}
\usepackage{graphicx,color}
\usepackage{amssymb}
\usepackage{amsmath}
\usepackage{xspace}
\usepackage{url}

\begin{document} %%%%%%%%%%%%%%%%%%%%%%%%%%%%%%%%%%%%%%%%%%%%%%%%%%%%%%%%%%
%------------------------------------------------------------------------------ 
% Title
%------------------------------------------------------------------------------
\title{Electron surface layer at the interface of a plasma and a dielectric wall}

%------------------------------------------------------------------------------ 
% Authors
%------------------------------------------------------------------------------
%------------------------------------------------------------------------------ 
% Date
%------------------------------------------------------------------------------
\author{R. L. Heinisch, F. X. Bronold, and 
H. Fehske}
\affiliation{Institut f{\"ur} Physik,
             Ernst-Moritz-Arndt-Universit{\"a}t Greifswald,
             17489 Greifswald,
             Germany}

\date{\today}
\begin{abstract}
We study the plasma-induced modifications of the potential and charge distribution across the interface of 
a plasma and a dielectric wall. For this purpose, the wall-bound surplus charge arising from the plasma is 
modelled as a quasi-stationary electron surface layer in thermal equilibrium with the wall. It satisfies 
Poisson's equation and minimizes the grand canonical potential of wall-thermalized excess electrons. Based 
on an effective model for a graded interface taking into account the image potential and the offset of the 
conduction band to the potential just outside the dielectric, we specifically calculate the modification
of the potential and the distribution of the surplus electrons for MgO, SiO$_2$ and Al$_2$O$_3$ surfaces in 
contact with a helium discharge. Depending on the electron affinity of the surface, we find two vastly different 
behaviors. For negative electron affinity, electrons do not penetrate into the wall and a quasi-two-dimensional 
electron gas is formed in the image potential, while for positive electron affinity, electrons penetrate into 
the wall and a negative space charge layer develops in the interior of the dielectric. We also investigate how 
the non-neutral electron surface layer -- which can be understood as the ultimate boundary of a bounded gas 
discharge -- merges with the neutral bulk of the dielectric.

\end{abstract}
\pacs{52.40.Hf, 73.30.+y, 52.80.Tn}
\maketitle

%\twocolumn

\section{Introduction}

Macroscopic objects in contact with an ionized gas acquire a negative charge because the 
influx of electrons from the plasma outruns the influx of ions. The collection of electrons 
at the wall (boundary of the object) gives rise to a repulsive Coulomb potential 
which reduces the electron influx until the wall charge reaches a quasi-stationary value. 
As a consequence of the electron accumulation at the wall an electron depleted region, the
plasma sheath, is formed adjacent to the wall. 

Most of the voltage driving the discharge drops across the sheath.
Wall charges may however not only affect the spatial structure of the plasma but also 
surface-supported elementary process such as electron-ion recombination and secondary 
electron emission, which are particularly important in dusty plasmas,\cite{Ishihara07,FIK05,Mendis02} 
dielectric barrier discharges,\cite{Kogelschatz03,SAB06,SLP07} and solid-state based 
micro-discharges.\cite{DOL10,WTE10,BSE06,OE05,Kushner05} A macroscopic description of the 
plasma-induced wall 
charge, sufficient for the modeling of the plasma sheath,~\cite{Franklin76} is clearly insufficient
for quantifying the effect wall charges might have on these processes. A microscopic description
of the plasma-induced wall charge and the potential across the plasma-wall interface it leads to is required.

Traditionally, plasma walls are treated as perfect absorbers.\cite{LL05,Franklin76,Riemann91} 
Irrespective of the microscopic interaction, all electrons and ions impinging on the wall are 
assumed to recombine instantly. From this model only the wall potential just outside the wall
can be obtained. This is the potential that balances the electron and ion influx at the wall. A first, 
qualitative step going beyond this model was taken by Emeleus and Coulter,\cite{EC87,EC88} who 
envisaged the wall charge as a two-dimensional surface plasma coupled to the bulk plasma via 
electron-ion wall recombination. No attempt was however made to put this appealing idea onto 
a formal basis. Later the notion of a two-dimensional surface charge was developed further by Behnke 
and coworkers\cite{BBD97,USB00,GMB02} utilizing phenomenological rate equations for the electron 
and ion surface densities. In these equations, the microphysics at the wall is encapsulated in 
surface parameters, such as, electron and ion sticking coefficients, electron and ion desorption 
times, and an electron-ion wall recombination coefficient. In principle these parameters 
can be calculated. Assuming, for instance, plasma electrons to ad- and desorb in the long-range 
image potential of the wall we calculated in our previous work electron sticking
coefficients and electron desorption times for uncharged metallic~\cite{BDF09} and dielectric 
surfaces.~\cite{HBF10a,HBF10b,BHMF11} We also made a first attempt to estimate these two quantities 
for charged dielectric plasma walls~\cite{HBF11} and proposed a physisorption-inspired microscopic 
charging model for dust particles in a gas discharge.~\cite{BFKD08}

In this work, we shift gears and focus on the potential and charge
distribution across the plasma-wall interface after the 
quasi-stationary wall charge (electron adsorbate in the notation of our previous
work~\cite{BDF09,HBF10a,HBF10b,BHMF11,HBF11,BFKD08}) has been established. In other words, 
we extend the modeling of the plasma sheath~\cite{LL05,Franklin76,Riemann91}
to the region inside the solid and calculate the plasma-induced modifications of the
potential and charge distribution of the surface. Although knowing the potential and charge 
distribution across the interface may not be of particular importance for present day 
technological plasmas, it is of fundamental interest from an interface physics
point of view. In addition, considering the plasma wall as an integral part of the plasma 
sheath may become critical when the miniaturization of solid-state based plasma 
devices~\cite{DOL10,WTE10,BSE06,OE05,Kushner05} continues.

In the model outlined below we specifically consider a dielectric wall and treat 
the plasma-induced quasi-stationary wall charge, that is, the surplus charge on top 
of the charge distribution of the bare, free-standing surface, as an electron surface 
layer of a certain extent, which is trapped by and in thermal equilibrium with the wall. 
In order to determine the chemical potential, width, and spatial position (relative to 
the crystallographic interface) of the electron surface layer, which depend on
surface as well as plasma parameters, we employ a one-dimensional model for a graded
interface between a collisionless plasma sheath and a dielectric surface which is assumed
to be a perfect absorber, that is, the wall potential balances at a certain distance from
the crystallographic interface the electron and the ion influx from the plasma.

The model of a graded interface encompasses two important ingredients: the surface dipole 
of the bare surface responsible for the offset of the conduction band minimum to the potential 
just outside the dielectric and the long-range image potential. The former accounts for the 
charge re-distribution of the free-standing, uncharged surface arising from the truncation of 
the crystal and the latter supports polarization-induced external surface states (image states),
first predicted for liquid helium~\cite{CC69} and later studied for metallic and dielectric 
surface with negative electron affinity~\cite{SZG09,SCE99,BKP07,RWK03,HSR97,Fauster94}, which 
may trap the electron surface layer in front of the crystallographic interface. 

Originally proposed by Stern for the interface between two 
dielectrics,\cite{Stern78} and later on used by others for semiconductor heterojunctions~\cite{SS84} 
and electron trapping in nanopores,\cite{PM06} the graded interface model also guarantees 
continuity of the electrostatic potential across the plasma-wall interface. The model thus allows 
us to study the spatial distribution of the plasma-induced wall charge across the interface. To insert 
the surplus charge into the interface we follow Tkharev and Danilyuk~\cite{TD85} and minimize, in 
the spirit of density functional theory,~\cite{Mermin65,KS65,JG89} the grand canonical potential of  
wall-thermalized excess electrons. We also investigate how the electron surface layer merges with 
the neutral bulk of the dielectric which we describe with the model of an intrinsic semiconductor.

Various improvements of the model are conceivable but the increased mathematical complexity 
would mask the general ideas we would like to convey. For instance, the model of a collisionless 
sheath could be replaced by more realistic models.~\cite{LL05,Franklin76,Riemann91} Going beyond 
the perfect absorber model, on the other hand, is an unsolved problem. It would require the inclusion 
of electron desorption, electron sticking, and electron-ion recombination, with the respective 
coefficients to be self-consistently calculated for the quasi-stationary electron adsorbate at the 
wall. Replacing the graded interface by an ab-initio theory for the surface, for instance, along 
the lines given in Refs.~\cite{GMG10,BKP07,RWK03,JJW88},  
possibly taking ad-layers of the host gas' atoms or molecules as well as impurities and 
imperfections into account, is desirable but at the present stage of the investigation impractical. 
It would require an expensive atomistic characterization of the plasma-wall interface, either 
experimentally via various surface diagnostics or theoretically via ab-initio simulations. As long 
as the atomistic details affect however only the off-sets of the dielectric constant, the electron 
affinity, and the effective mass, the graded interface model already incorporates these details by a 
suitable parameterization. What is not well described is the non-universal region a few atomic units 
below and above the crystallographic interface. In particular, intrinsic surface states (Shockley 
and Tamm states~\cite{Lueth92}) and additional surface states which may arise from the short-range 
surface potential due to impurities, imperfections, and ad-layers 
are not included. If unoccupied these states could trap the electron surface layer in the 
vicinity of the interface even for surfaces with positive electron affinity where image 
states are absent and cannot trap the surplus charge in front of the 
surface.

The remaining paper is structured as follows. In Sec.~\ref{Electron surface layermain} we first construct 
a crude model for the plasma-induced electron surface layer at the interface between a plasma and a 
dielectric wall. It does not account for the merging of the electron surface layer with the bulk of the 
dielectric. As long as the primary interest is in the region close to the crystallographic interface and 
the band gap of the dielectric is large enough, the crude model is sufficient. Section~\ref{NegativeSCR} 
describes a refinement of the model which enables one to also investigate the cross-over of the 
electron surface layer to the bulk of the dielectric. This is particularly important for dielectrics 
with small energy gaps. Numerical results for the potential and the electron distribution are given in 
Section \ref{Results} and a short summary is formulated in Sec. \ref{Summary}.

\begin{figure}
\includegraphics[width=\linewidth]{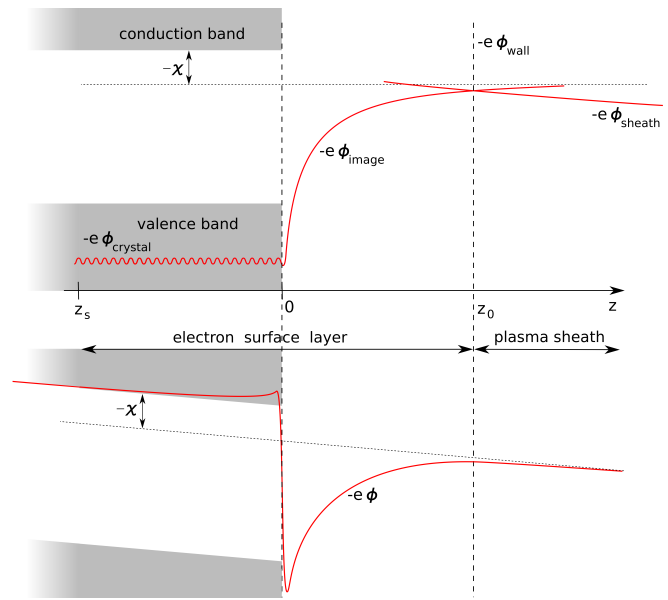}
\caption{Qualitative sketch of an interface between a plasma and a dielectric wall. Upper panel: 
Band structure, microscopic crystal potential merging with the image potential, and sheath 
potential. Lower panel: Effective potential for the graded interface on which the model of an 
electron surface layer is based.}
\label{figure1}
\end{figure}

\section{Crude electron surface layer}
\label{Electron surface layermain}

As depicted in Fig.~\ref{figure1}, we consider an ideal, planar interface at \(z=0\) with the dielectric 
occupying the half-space $z<0$ and the discharge occupying the half-space $z>0$. Chemical contamination
and structural damage due to the burning gas discharge are discarded. At the moment we focus on the 
physical principles controlling the electronic properties of the plasma-wall interface. In the model we 
propose the plasma-induced wall charge to be treated as an electron surface layer (ESL) which is an 
interface specific electron distribution on top of the charge re-distribution due to the truncation 
of the solid. The ESL is assumed to be thermalized with the solid and to stretch from the plasma sheath 
over the crystallographic interface to the bulk of the dielectric.

The boundary between the ESL and the plasma sheath is located in front of the surface at \(z=z_0\). It 
is the position where the attractive force due to the surface potential \(\phi_\text{\it surf}\) and the 
repulsive force due to the sheath potential \(\phi_{sheath}\) balance each other. Thus, \(z_0\) is given by 
\begin{align}
\phi_\text{\it surf}^\prime(z_0)+\phi_{sheath}^\prime(z_0)=0 \text{ .} \label{z0condition}
\end{align}
It gives the position of an effective wall for plasma electrons and ions at which, for instance, 
the flux balance condition of the perfect absorber model, \(j_e=j_i\), with \(j_e\) and \(j_i\),
respectively, the electron and ion flux towards the dielectric surface, has to be fulfilled. 
For \(z<z_0\) an electron is attracted to the surface and thus contained in the ESL while for \(z>z_0\) 
it is repelled back into the plasma. On the solid side, for \(z<0\), the ESL is bounded because 
of the shallow potential well formed by the restoring force from the positive charge in the plasma sheath. 

In this section we will outline the essential building blocks 
of the ESL model. Putting together concepts from plasma as well as surface physics a 
detailed, self-contained account seems to be helpful.

\subsection{Plasma sheath}

In the traditional view, electrons missing in the positive space charge region in
front of the plasma wall accumulate on the wall and give rise to a wall potential. For the construction
of our one-dimensional interface model we need the total number per unit area of missing sheath electrons (that 
is, the total surface density of missing sheath electrons) as a function of the wall potential because it 
is this number of electrons which can be distributed across the ESL. Hence, we require a model for the 
plasma sheath.

For simplicity, we use a collisionless sheath~\cite{Franklin76}, more realistic sheath 
models~\cite{LL05,Franklin76,Riemann91} make no difference in principle. 
In the collisionless sheath electrons are thermalized, 
that is, the electron density \(n_e=n_0 \exp(e \phi/k_B T_e) \), with \(\phi\) the potential, 
\(n_0\) the plasma density and \(T_e\) the electron temperature. The ions enter the sheath 
with a directed velocity \(v_{i0}\) and satisfy a source-free continuity equation, 
\(\mathrm{d(n_i v_i)}/\mathrm{d}z=0\), implying \(n_i v_i =n_0 v_{i0}\), and an equation of 
motion \(M(v_i \frac{\mathrm{d}}{\mathrm{d}z}v_i)=-e\frac{\mathrm{d}}{\mathrm{d}z} \phi\), with 
\(n_i\) the ion density, and \(M\) the ion mass. The potential \(\phi\) satisfies Poisson's equation 
\(\mathrm{d}^2\phi/\mathrm{d}z^2 =-4\pi e (n_i-n_e)\). Thus, the governing equations for the 
collisionless plasma sheath are~\cite{Franklin76}
\begin{align}
v_i \frac{\mathrm{d}v_i}{\mathrm{d}z}&=-\frac{e}{M} \frac{\mathrm{d} \phi}{\mathrm{d}z} \quad \text{ and} \label{sheath1}\\
\frac{\mathrm{d}^2}{\mathrm{d}z^2}\phi &=-4\pi e n_0 \left[ \frac{v_0}{v_i}-\exp \left( \frac{e \phi}{k_B T_e} \right) \right] \text{ .} \label{sheath2}
\end{align}
Using dimensionless variables
\begin{align}
\eta=-\frac{e\phi}{k_BT_e} \text{ , } \quad  \xi=\frac{z}{\lambda_D} \text{ , } \quad \text{and} \quad u=\frac{v_i}{c_s}
\end{align}
where 
\begin{align}
 \lambda_D=\sqrt{\frac{k_B T_e}{4 \pi n_0 e^2}} \quad \text{and}\quad c_s=\sqrt{\frac{k_BT_e}{M}} 
\end{align}
equations (\ref{sheath1}) and (\ref{sheath2}) become
\begin{align}
u u^\prime &= \eta^\prime \quad \text{ and} \\
\eta^{\prime \prime}&= \frac{u_0}{u} - \exp(-\eta) \text{ .}
\end{align}

In the ESL model the plasma occupies not the whole half space \(z>0\) but only the portion
\(z>z_0\) (see Fig.\,\ref{figure1}). The 
integration of the first equation 
gives \mbox{\(u=-\sqrt{2\eta+u_0^2}\)}, where \(u_0=v_{i0}/c_s\) is the reduced velocity 
of ions entering the sheath, so that the second equation becomes
\begin{align}
\eta^{\prime \prime}=-\frac{u_0}{\sqrt{2 \eta +u_0^2}}-\exp(-\eta) \text{ .} \label{clsheatheq}
\end{align}
Using the boundary condition that the potential and the field vanish far inside the plasma, that is, 
\(\eta \rightarrow 0\) and \(\eta^\prime \rightarrow 0\) for \(\xi\rightarrow\infty\), Eq. (\ref{clsheatheq}) 
can be integrated once and we obtain
\begin{align}
\eta^\prime=-\sqrt{-2 u_0 \sqrt{2 \eta +u_0^2}+2 \exp (-\eta) +2 u_0 \sqrt{u_0^2}-2} \text{ .} 
%\label{etaprime}
\end{align}
For ions entering the sheath with the Bohm velocity \(u_0=-1\). The field at the wall 
as a function of the wall potential \(\eta_w=\eta(\xi_0)\) is then given by 
\begin{align}
\eta_w^\prime=-\sqrt{2\sqrt{2\eta_w+1}+2\exp(-\eta_w)-4} \text{ .} \label{etawprime}
\end{align}

The total surface density of electrons in the ESL equals the total surface density of missing sheath 
electrons, in other words, the total surplus surface density of positive ions in the sheath 
\(N\) which can be calculated from the electric field at the wall. Integrating Poisson's equation yields 
\begin{align}
N&=\int_{z_0}^\infty \mathrm{d}z (n_i-n_e) =- \frac{1}{4 \pi e} \int_{z_0}^\infty \mathrm{d}z \frac{\mathrm{d}^2\phi}{\mathrm{d}z^2} \nonumber \\
&=\frac{1}{4\pi e} \frac{\mathrm{d}\phi}{\mathrm{d}z}(z_0)=-n_0 \lambda_D \eta_w^\prime \text{ .} \label{Neq} 
\end{align}
Combing Eqs.~(\ref{Neq}) and (\ref{etawprime}) gives the total surface density of electrons to be inserted into 
the ESL as a function of the wall potential.

The wall potential itself is determined by the flux balance condition, $j_e=j_i$, which, in the ESL model, 
is assumed to be fulfilled at \(z=z_0\). Using the Bohm flux for the ions and the thermal flux for the electrons,
\begin{align}
j_i=n_0 \sqrt{\frac{k_B T_e}{M}}\quad  \text{and} \quad j_e=\frac{1}{4} n_0 \sqrt{\frac{8 k_B T}{\pi m_e}} e^\frac{e\phi}{k_B T_e} \text{ ,}
\end{align}
the wall potential is given by~\cite{Franklin76}
\begin{align}
\eta_w=\frac{1}{2} \ln \left( \frac{M}{2\pi m_e} \right)  \text{, }
\end{align}
that is,
\begin{align}
 \phi_w=-\frac{k_B T_e}{2 e} \ln\left(\frac{M}{2 \pi m_e}\right) \text{ .}
\end{align}
In the collisionless sheath model the wall potential depends only on the electron temperature and 
the ion to electron mass ratio. 

\subsection{Surface dipole}
We now turn to the interface region in which the missing sheath electrons will be inserted. This region
is absent in the traditional modeling of plasma walls. In our model it is an extended region surrounding
an ideal dielectric surface. In comparison to the electrons responsible for the chemical binding within
the dielectric the additional electrons coming from the plasma are only a few. The electronic structure 
of the surface, in particular, the charge re-distribution due to truncation of the solid and the offset
of the energy bands in the bulk with respect to the potential outside the dielectric will not be changed 
significantly by the presence of the surplus electrons comprising the wall charge. 
 
In order to quantify the above statement let us first consider the electrostatic potential and the 
electronic structure of a free-standing, uncharged dielectric surface. According 
to Tung,\cite{Tung01} it has to minimize the thermodynamical potential and satisfy Poisson's equation
implying that the potential is continuous across the surface. Strictly speaking, the continuity of the 
potential only applies to the microscopic crystal potential which has to merge continuously with the
surface potential outside the crystal. The averaged long-range potential, in contrast, can be  
discontinuous at the interface. It is this offset which is encoded in the surface dipole.

The energy of an electronic state in the bulk of the dielectric can be referenced to the vacuum level 
\(V(\infty)=0\), that is, the potential far outside the crystal, in the following way,\cite{Tung01} 
\begin{align}
E_{ik}(\vec{r})=\epsilon_{ik}-e\bar{V}_{cell}-eV_{s}(\vec{r}) \text{ ,} \label{energyreference}
\end{align}
where \(\epsilon_{ik}\) is the quantum-mechanical contribution to the energy, \(\bar{V}_{cell}\) 
is the averaged potential of a cell due to the charge distribution within the same cell, and \(V_s(\vec{r})\)
is the long range potential due to the surface dipole, space charges, and 
external fields. In the simple two-band model depicted in Fig.~\ref{figure1}, \(i=v,c\). \(V_{s}(\vec{r})\)
contains the surface dipole arising from the truncation of the solid and responsible for the potential offset 
at the surface and a slowly varying component due to external fields and internal and plasma-induced space 
charges. External fields and internal space charges will 
be neglected in the following and plasma-induced space charges will be accounted for by Poisson's equation
(see below). 

In order to judge whether the surplus charge arising from the plasma affects the surface dipole 
it is useful to consider first the typical strength of the surface dipole of a free-standing, uncharged 
dielectric surface. It results from a charge double-layer in immediate proximity to the
surface. Depending on the material it can have various origins. For an ionic crystal, for example, it is the 
lattice relaxation at the surface which makes anions or cations to protrude and the other species to retract 
(e.g. protruding oxygen and retracted cations for magnesium oxide\cite{BKP07}), while for semiconductors it 
is the regrouping of covalent bonds which leads to charge re-distribution at the surface. Even in the
absence of these effects the minimization of the thermodynamic potential of the surface's electrons leads 
already to an electron density leaking out into the vacuum. This is particularly important for metals.
As a result a charge double-layer is formed over a length on the order of a lattice constant. 

The dipole layer is usually characterized by a dipole strength
\begin{align}
eD=eV_s(\vec{r}_s^-)-eV(\vec{r}_s^+) \text{ ,}
\end{align}
where \(V_s(\vec{r}_s^-)\) is the limit of the long range potential just inside the crystal at the surface 
position \(\vec{r}_s\) and \(V(\vec{r}_s^+)\) is the limit of the potential at that position just outside the 
surface. Usually these two potentials, which characterize the discontinuity of the long-range potential
at the surface, are termed the potential just inside and the potential just outside, 
respectively.\cite{Tung01,CK03} Here, just outside denotes a distance small compared to variations of the 
long-range potential due to external fields or space charges but large compared with the width of the charge 
double-layer. Note also that in the definition of the potential just outside the image potential is assumed 
to have already decreased to zero.~\cite{CK03} This point will be important later.

The strength of the dipole layer is a microscopic property of the surface which is relatively insensitive 
against the additional charges from the plasma. The reason for this lies in the small number of additional 
electrons from the plasma compared to the number of displaced electrons involved in the formation of the dipole 
layer. To prove this statement we give a simple estimate. Typical surface dipoles \(eD\) are on the order of 
electron Volts. For a double-layer of one \(\mathring{A}\) to each side of the crystallographic interface a potential 
difference of one Volt requires a surface charge density of \(5.5 \times 10^{13} cm^{-2}\). The surface charge 
density at the wall of a helium discharge with plasma density \(n_0=10^{7} cm^{-3}\) and electron temperature 
\(k_BT_e=2eV\) amounts, however, only to \(4.4 \times 10^{6} cm^{-2}\). For typical plasma densities, the number 
of additional electrons is thus far too small to lead to a change of the surface dipole. Even for 
semiconductor-based microdischarges~\cite{WTE10,OE05}, which can have much higher plasma densities,
we expect the surface dipole of the plasma wall not to be modified by the plasma.
 
In view of the just given estimate, we have to revise an assumption in our previous work,\cite{BHMF11} where we
assumed the surface charge accumulating on the wall would increase the dipole energy \(eD\) by 
\(e\phi_w\), leading to the image states being pushed from the band gap into the energy region of the conduction band. 
The numbers given in the previous paragraph indicate, however, that the band line-up of the conduction band and the 
potential just outside the solid, will not be affected much by the wall charge. Hence, if a negative electron affinity 
supports image states in front of the uncharged surface, these states remain in the band gap for the charged 
surface. Electron trapping as investigated in Refs.~\cite{HBF10a,HBF10b} is thus even possible for charged 
plasma walls.

Instead of the dipole strength \(eD\) which cannot be measured directly, it is more convenient to characterize the 
dipole layer by the electron affinity \(\chi\) which is a measurable quantity for a dielectric surface.~\cite{CK03} 
The electron affinity is the energy released when an electron is moved from just outside the surface to the bottom 
of the conduction band. It accounts for charge re-distribution in the vicinity of the surface due to the
truncation of the crystal. While many surfaces have positive electron affinity
such as Al$_2$O$_3$  or SiO$_2$, there are also materials with negative electron affinity, for instance, 
diamond~\cite{HKV79}, boron nitride\cite{LSG99}, or the alkaline earth oxides.~\cite{BKP07,RWK03}
The electron affinity depends also on ad-atoms. In some cases this is even used to control the electron 
affinity of a surface. Terminating, for instance, a surface with weakly electronegative elements such as 
hydrogen induces a negative electron affinity,~\cite{CRL98} while termination with strongly electronegative 
elements can lead to a positive electron affinity.~\cite{FRL01} 

From Eq. (\ref{energyreference}) it is clear that \(\chi\) equals \(eD\) plus a bulk contribution, 
\begin{align}
\chi=-eV_s(\vec{r}_s^+)-E_{ck}(\vec{r}_s^-)=eD-E_{C}+e\bar{V}_{cell}~,
\end{align}
where \(E_C=\epsilon_{cm}\) denotes the minimum of the conduction band. We can thus 
use \(\chi\) to characterize the potential offset at the surface. There is however a caveat. The long-range 
potential inside the solid is only specified up to a constant.\cite{Tung01} Typical choices are the cell-averaged 
potential or the inter-sphere potential of the muffin-tin approximation. For our purpose it will be however more 
convenient to take the conduction band minimum as the long-range potential inside the solid. This choice is motivated 
as follows. We are considering a dielectric with a large energy gap. The valence band is thus fully occupied and the 
conduction band is essentially empty. Hence, only the conduction band can be populated by additional electrons 
coming from the plasma and referencing the electrostatic potential inside the solid to the conduction band minimum 
allows us to relate the total surplus electron density in the interface region to the potential in the interface 
region in analogy to what we have done for the plasma sheath in the previous subsection. 

Adopting the above discussion to the one-dimensional model shown in Fig.~\ref{figure1} and assuming a quadratic 
dispersion for the conduction band, the energy of an electron in the conduction band is given by
\begin{align}
E_k(z)=\frac{\hbar^2 k^2}{2m_C^*}-e\phi_{surf}(z)~, 
\end{align}
where \(\phi_{surf}(z)\) is the total surface potential to be determined in the next subsection and the offset 
of the electrostatic potential at the surface,
\begin{align}
e\phi_{surf}(0^-)-e\phi_{surf}(0^+)=\chi~,
\label{TotalPhi}
\end{align}
encompasses the surface dipole as well as the unspecified bulk contribution.

%The electron affinity \(\chi\) encompasses therefore the potential offset 
%at the crystallographic interface over the length scale of the lattice constant. 
%The band structure and the surface potential at the surface of a dielectric are visualized in Fig. \ref{figure1}.
 
\subsection{Image potential}

The surface potential of the bare, uncharged surface comprises at least the surface dipole and a long-range 
contribution, the image potential, resulting from the mismatch of the dielectric constants at the surface. 
Far away from the surface the image potential is given by\cite{Jackson98} 
\begin{align}
\phi_{im}(z)=\frac{\epsilon-1}{4(\epsilon+1)}\frac{e}{z} \text{ .}
\label{ImageClassical}
\end{align}
But this expression cannot be employed for our purpose because the singularity at \(z=0\) prohibits 
a smooth electron distribution across the interface. In reality the image potential has to continuously 
merge with the crystal potential. Equation~(\ref{ImageClassical}) is thus also unphysical.

To obtain a realistic image potential without performing an atomistically accurate calculation we employ  
the model of a graded interface. It also has the virtue to be parameterizable with experimentally measured 
values for the electron affinity, the dielectric mismatch, and the mismatch between effective electron 
masses. The model incorporates therefore important properties of a surface, most importantly, it accounts 
for the charge re-distribution due to the truncation of the solid. 

Initially proposed by Stern~\cite{Stern78} to remove the unphysical singularity of the image potential at the 
interface of two dielectrics, the graded interface model assumes the dielectric constant \(\epsilon\) to vary 
over a distance on the order of a lattice constant. Later the model was extended to variations of other 
physical quantities and applied to semiconductor heterostructures and nanopores.~\cite{SS84,PM06} Clearly,
because of the interpolation the model cannot account for effects associated with intrinsic surface states 
(Shockley and Tamm states~\cite{Lueth92}) and additional surface states which may arise from the short-range 
surface potential. Nevertheless, the graded interface model is a reasonable description of a surface.
 
In the spirit of a graded interface, we assume the dielectric constant \(\epsilon\), the electron mass
\(m\), and the potential offset at the surface to vary smoothly according to the grading 
function
\begin{align}
g_{c^-,c^+}(z)=\left\lbrace \begin{matrix}
c^- & z<-a \\
\frac{c^-+c^+}{2}-\frac{c^--c^+}{2}\sin \left( \frac{\pi x}{2 a}\right) &-a<z<a~, \\
c^+ & z>a
\end{matrix} \right.
\end{align} 
where \(a\) is the half width of the graded interface and \(c^\mp\) stands for the quantity that varies across the 
interface. We use the value \(a=5 \mathring{A}\) which is an estimate used in previous applications of the graded 
interface.\cite{Stern78,SS84,PM06} While the grading parameter \(a\) is not based on definite experimental or 
theoretical results it is motivated by the assumption that the bonding and the electron density at the surface change 
over one to two lattice constants implying a transition layer for the effective potential and the dielectric constant 
that is somewhat larger. Hence, across the interface the electron mass, the dielectric constant and the offset potential
are given by
\begin{align}
m(z)=g_{m^\ast_C,m_e}(z) \text{ , } \quad \epsilon(z)=g_{\epsilon,1}(z)  \text{ , } \label{graded1}
\end{align}
and
\begin{align}
\phi_{\text{\it offset}}(z)=\frac{1}{e}g_{\chi,0}(z) \text{ ,} \label{gradedoffsetpotential}
\end{align}
respectively, with \(m_C^\ast\) the effective mass of the conduction band.

Within the model of the graded interface, the image potential is the change in the selfenergy 
of an electron due to the proximity of the dielectric mismatch. Positioning the electron at \(\vec{r}_0\) 
it is given by~\cite{Stern78}
\begin{align}
\phi_{im}(\vec{r}_0)=\frac{1}{2}\left[\phi^m(\vec{r}_0)-\phi^0(\vec{r}_0) \right]~,
\end{align}
where \(\phi^m(\vec{r})\) is the potential in the medium with dielectric mismatch arising from the 
electron at \(\vec{r}_0\) and \(\phi^0(\vec{r})\) is the same quantity in a homogeneous medium with 
dielectric constant \(\epsilon(z_0)\). 
Hence, \(\phi^m(\vec{r})\) is the solution of
\begin{align}
\nabla \left( \epsilon(z) \nabla \phi^m(\vec{r}) \right) = 4 \pi e \delta \left( \vec{r}- \vec{r}_0 \right) \label{potmismatch}
\end{align}
while \(\phi^0(\vec{r})\) is the solution of
\begin{align}
\nabla^2 \phi^0(\vec{r})=\frac{4 \pi e}{\epsilon(z_0)}\delta(\vec{r}-\vec{r}_0) \text{ .} \label{potnomismatch}
\end{align}

To solve Eqs. (\ref{potmismatch}) and (\ref{potnomismatch}) we follow Stern\cite{Stern78} and make the ansatz
\begin{align}
\phi^{m,0}(z,\rho,\phi)=\frac{1}{2\pi}\sum_{l=-\infty}^\infty & \int_0^\infty  \mathrm{d}q q J_l(\rho q) J_l(\rho_0 q) \nonumber \\ &\times e^{il(\phi-\phi_0)}A_q^{m,0}(z) \text{ .}
\end{align}
Placing the electron on the \(z\) axis, \(\rho_0=0\) which implies \(J_0(\rho_0 q)=1\) and \(J_l(\rho_0 q)=0\) for \(l>0\). Hence, we need to keep only the \(l=0\) term, so that
\begin{align}
\phi^{m,0}(z,\rho,\phi)=\frac{1}{2\pi} \int_0^\infty \mathrm{d}q q J_0(\rho q)  A^{m,0}_q(z) \text{ ,}
\end{align}
where \(A_q^0(z)\) is given by
\begin{align}
A_q^0(z)=\frac{-2\pi e}{ \epsilon(z_0)q}e^{-q|z-z_0|}
\end{align}
and \(A^m_q(z)\) is the solution of
\begin{align}
A_q^{m\prime \prime}(z)+\frac{\epsilon^\prime(z)}{\epsilon(z)} A_q^{m \prime}(z) -q^2A^m_q(z)=\frac{4 \pi e}{\epsilon(z)} \delta(z-z_0)
\end{align} 
which has to be obtained numerically. The image potential is then given by
\begin{align}
\phi_{im}(z_0)=\frac{1}{4 \pi} \int_0^\infty \mathrm{d}q q\left(A^m_q(z_0)-A_q^0(z_0) \right) \text{ .}
\end{align}
In contrast to Eq.~(\ref{ImageClassical}) it is now smoothly varying across the interface with a deep well 
on the low-\(\epsilon\) side and a small bump on the high-\(\epsilon\) side. 

The total surface potential comprises the graded offset potential (\ref{gradedoffsetpotential}) 
and the graded image potential. Hence, 
\begin{align}
\phi_{surf}(z)=\phi_{im}(z)+\phi_{\text{\it offset}}(z) \text{ .} \label{empty_surf_pot}
\end{align} 
It is continuous across the crystallographic interface at \(z=0\) and enables us thereby 
to also calculate a smoothly varying electron distribution in the ESL. The band structure
and the total surface potential at the graded interface are visualized in the lower panel of Fig. \ref{figure1}.

Using Eq.~(\ref{z0condition}) we can now determine the position \(z_0\) of the effective wall, that is, 
the maximum extent of the ESL on the plasma side. The derivative of the bare surface potential is 
\(\phi_{surf}^\prime =\phi_{\text{\it offset}}^\prime+\phi_{im}^\prime\).  Due to the
relatively weak field in the sheath compared to the image force, the boundary \(z_0\) will be so far away 
from the interface that \( \phi_{\text{\it offset}}^\prime\) vanishes and the image potential 
obeys~(\ref{ImageClassical}). Thus, the boundary between the ESL and the 
plasma sheath is given by 
\begin{align}
z_0=\sqrt{\frac{(\epsilon -1) e}{4 (\epsilon +1) \phi_{w}^\prime}} 
\end{align}
with \(\phi_w^\prime=-(k_BT_e \eta_w^\prime)/(e\lambda_D)\) and \(\eta^\prime_w\) given by~(\ref{etawprime}).

\subsection{Electron distribution}
\label{Electron surface layer}

The plasma-induced wall charge is assumed to be in thermal equilibrium with the wall. Hence, the distribution 
of the excess electrons in the ESL has to minimize the excess electron's 
grand canonical potential in the external potential due to the surface. The coupling to the sheath is 
maintained by the constraint that only as many electrons can be filled into the ESL as are missing in
the sheath and the boundary conditions to the Poisson equation which links the electron distribution 
in the ESL to the (internal) electrostatic potential. 

To minimize the grand canonical potential of the surplus electrons we follow Tkharev and Danilyuk~\cite{TD85} 
and apply density functional theory~\cite{KS65,Mermin65} to the graded interface. While more refined schemes of density 
functional theory\cite{JG89} could, in principle, be employed, we will use for the purpose of this exploratory 
calculation  density functional theory in the local approximation.  Quite generally, the grand canonical potential
of an electron system in an external potential \(V(\vec{r})\) is given in the local approximation by
\begin{align}
\Omega=&\int V(\vec{r}) n(\vec{r}) \mathrm{d}\vec{r} -\frac{e}{2} \int \phi_C(\vec{r}) n(\vec{r}) \mathrm{d}\vec{r} \nonumber \\
& +G[n]- \mu \int n(\vec{r}) \mathrm{d}\vec{r} \text{ ,}
\end{align}
where \(G[n]\) is the grand canonical potential of the homogeneous system with density \(n(\vec{r})\) and the Coulomb potential is determined by
\begin{align}
\nabla \left( \epsilon(\vec{r}) \nabla \phi_C(\vec{r})\right)=4\pi e n(\vec{r}).
\end{align}
The ground state electron density minimizes \(\Omega\), that is, it satisfies
\begin{align}
V(\vec{r})-e\phi_C(\vec{r}) + \mu^h(n)-\mu=0~,
\label{DFT}
 \end{align}
where \(\mu^h(n)=\delta G[n]/\delta n\) is the chemical potential for the homogeneous system. 
 
Specifically for the excess electrons in the one-dimensional graded interface Eq.~(\ref{DFT}) reduces to 
\begin{align}
-e\phi(z)+\mu^h(z)-\mu=0 \text{ ,} \label{vareq}
\end{align}
where \(\mu^h(z)\equiv\mu^h(n(z),T)\) and the electrostatic potential, 
\begin{align}
\phi(z)=\phi_{surf}(z)+\phi_C(z) \text{ ,} \label{potDFT}
\end{align}
consists of the potential of the bare surface given by Eq.~(\ref{TotalPhi}) and the internal Coulomb potential 
which satisfies Poisson's equation,
\begin{align}
\frac{\mathrm{d}}{\mathrm{d}z}\left( \epsilon(z) \frac{\mathrm{d}}{\mathrm{d}z}\phi_C(z) \right)=4\pi e n(z) \text{ ,} \label{poiseq}
\end{align}
with the graded dielectric constant \(\epsilon(z)\) given by Eq. (\ref{graded1}) and the boundary conditions 
\(\phi_C(z_0)=\phi_w\) and \(\phi_C^\prime(z_0)=\phi_w^\prime\) to guarantee continuity of the potential at 
\(z_0\) and to include the restoring force from the positive charge in the sheath. Note that the Coulomb potential 
derived from this equation includes the attraction of an electron to the image of the charge distribution.

For the functional relation \(\mu^h(z)\equiv\mu^h(n(z),T)\) we take the expression adequate for a homogeneous, 
non-interacting, non-degenerate electron gas, 
\begin{align}
n(z)=\frac{1}{\sqrt{2}}\left( \frac{m(z) k T}{\pi \hbar^2}\right)^\frac{3}{2} e^\frac{\mu^h(z)}{k_BT} \text{ .} \label{neq}
\end{align}
This is justified because the density of the excess electrons is rather low and the temperature of the surface is 
rather high, typically a few hundred Kelvins. 

In order to calculate the quasi-stationary distribution of the surplus electrons, Eqs.~(\ref{vareq}) and (\ref{poiseq}) 
have to be solved self-consistently with the additional constraint that the total electron surface density in the ESL 
equals the total surface density of electrons missing in the sheath, that is, 
\begin{align}
\int_{z_s}^{z_0} \mathrm{d}z n(z)= N \text{ ,}
\label{Nconstraint}
\end{align}
with \(N\) given by Eq. (\ref{Neq}). In the above equation we introduced a cut-off \(z_s<0\) at which the 
ESL terminates inside the dielectric. As long as \(|z_s|\) is chosen large enough it does not affect the
numerical results close to the surface. An improved treatment of the ESL, avoiding the ad-hoc cut-off,
is given in the next section.

Within the crude ESL model developed in this section the computation is performed iteratively in the interval 
\(z_s<z<z_0\) according to the following scheme:
(i) We start with the potential of the empty surface given by Eq.~(\ref{potDFT}) with \(\phi_C(z)\)  
obtained from Eq.~(\ref{poiseq}) with \(n(z)=0\) but with the boundary conditions at \(z_0\) as specified. (ii) 
We integrate both sides of Eq.~(\ref{neq}) over \(z\) with \(\mu^h(z)\) given by Eq.~(\ref{vareq}). Enforcing 
the constraint (\ref{Nconstraint}) determines \(\mu\). (iii) Using \(\mu\) we calculate from Eq.~(\ref{vareq}) 
a new \(\mu^h(z)\) which gives with Eq.~(\ref{neq}) a new electron density \(n(z)\). (iv) Lastly, we determine 
from  Eq.~(\ref{poiseq}) the electrostatic potential associated with the up-dated \(n(z)\). Steps 
(ii) - (iv) are iterated until \(\mu\), which is far below the conduction band edge because of the non-degeneracy of
the excess electrons, converges.  

\section{Refined electron surface layer}
%\subsection{Negative space charge in the dielectric}
\label{NegativeSCR}

In the previous section we have taken into account only the electron concentration in the conduction 
band of the dielectric due to the electrons coming from the plasma. For wide band gap materials this is 
justified, especially near the surface, as their concentration is much larger than the intrinsic carrier 
concentration. Deep inside the dielectric, however, charge neutrality is not enabled by a vanishing electron 
density, but by the electron density decreasing to its intrinsic value, which is then balanced by the 
intrinsic hole concentration in the valence band. 

To take this effect into account, which is particularly important when the additional electrons accumulate 
deep inside the bulk of the dielectric, we divide the ESL into two regions: a very narrow interface-specific 
region (ISR) and a wide space charge region (SCR) in the bulk of the dielectric. The parameter \(z_s\) denotes 
now no longer an ad-hoc cut-off but the boundary between the two regions. It has to be chosen so that the ISR 
includes the major effect of the image potential in the dielectric implying \(z_s < - z_0\). The electron 
distribution and the potential in the ISR are calculated using the density functional approach outlined in the 
previous section. In the SCR we use for simplicity the model of an intrinsic semiconductor to describe electron 
and hole densities as well as the long-range potential. As the energy bands in the dielectric follow the long-range 
potential the refined ESL also captures the band bending which might be induced by the presence of the wall
charge. It is however only significant when most of the excess electrons are trapped in the SCR and not in 
the ISR.
\begin{figure}
\includegraphics[width=\linewidth]{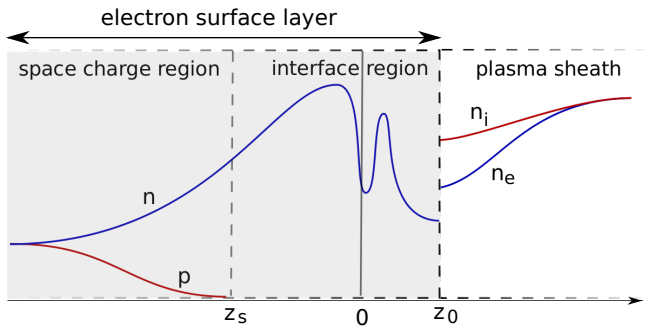}
\caption{Sketch of the refined model of the interface between a plasma and a dielectric wall. In the plasma 
equal densities of electrons and ions ensure quasi-neutrality. The positive space charge in front of the 
effective wall defines the plasma sheath. The ESL contains a very narrow interface-specific region (ISR) 
where the model of the graded interface is used and a wide space charge region (SCR) which allows a continuous 
merging with the neutral bulk of the dielectric where intrinsic electrons and holes balance each other
to guarantee charge neutrality. Note, the widths of the various regions are not to scale.}
\label{figure1a}
\end{figure}

Figure~\ref{figure1a} schematically shows the electron and hole densities for the refined ESL model. The 
boundary between the plasma sheath and the ESL is still located at \(z_0\). As our model does not encompass the 
electron and ion flux from the plasma for \(z<z_0\), the densities \(n_e\) and \(n_i\) are discontinuous at 
\(z_0\). This is obvious for the ions which are not allowed to enter the solid. The discontinuity of the electron 
density, in contrast, arises because we ensure only the total number of missing sheath electrons per unit area 
to be conserved. This global constraint cannot guarantee continuity of the electron density at \(z_0\). 
%Hot electrons on the plasma side which are reflected back and not thermalized within the solid are for instance 
%absent in the ESL. 
At the boundary between the ISR (\(z_s<z<z_0\)) and the SCR (\(z<z_s\)) the electron density and the potential 
are continuous. In principle, also the hole density \(p\) should be continuous. As \(p(z_s) \ll n(z_s)\) for the 
materials we are considering, we can however neglect holes in the ISR.
 
For the modeling of the SCR it is convenient to use \(\psi(z)=\phi(z)-\phi_{\text{bulk}}\) for the long range 
potential, which vanishes for charge neutrality in the bulk. Here, \(\phi_\text{bulk}= \phi( -\infty)\) (see 
below for an explicit relation for \(\phi_\text{bulk}\)).  Then, Poisson's equation is given by,
\begin{align}
\frac{\mathrm{d}^2\psi(z)}{\mathrm{d}z^2}=-\frac{4 \pi}{\epsilon}\left(-e n(z)+ep(z) \right) \text{ ,}
\end{align}
where the electron and hole densities for an intrinsic semiconductor with parabolic bands whose 
extremal points are, respectively, \(E_C\) and \(E_V\) are given by~\cite{AM76}
\begin{align}
n(z)&=\frac{1}{\sqrt{2}}\left( \frac{m_C^\ast k_BT}{\pi \hbar^2} \right)^\frac{3}{2} e^{\frac{1}{k_BT} (\nu-E_C+e\psi(z))}  \text{ ,} \label{compneq}\\
p(z)&=\frac{1}{\sqrt{2}}\left( \frac{m_V^\ast k_BT}{\pi \hbar^2} \right)^\frac{3}{2} e^{-\frac{1}{k_BT} (\nu-E_V+e\psi(z))} \text{ .}
\end{align}

From a comparison of the exponents in Eq. (\ref{compneq}) and Eq. (\ref{neq}), where \(\mu_h\) is given by Eq. 
(\ref{vareq}) we find
\begin{align}
\nu=\mu+E_C+e\phi_{bulk} \text{ .}
\end{align}
Far from the surface, \(\psi=0\) and \(n=p=n_b\). This gives the chemical potential
 \begin{align}
 \nu=\frac{E_V+E_C}{2} + \frac{3}{4} k_B T \ln \left(\frac{m_V^\ast}{m_C^\ast} \right)  
 \end{align}
and the bulk carrier concentration \(n_b\)
\begin{align} 
n_b=\frac{1}{\sqrt{2}}\left( \frac{k_B T}{\pi \hbar^2} \right)^\frac{3}{2} \left( m_C^\ast m_V^\ast \right)^\frac{3}{4} \exp \left(- \frac{E_{g}}{2k_BT} \right) \text{ ,}
\end{align}
where \(E_{g}=E_C-E_V\). Hence, Poisson's equation becomes
\begin{align}
\frac{\mathrm{d}^2 \psi(z)}{\mathrm{d}z^2}=\frac{4 \pi e}{\epsilon} n_b \left( e^{\frac{ e \psi(z)}{k_BT}}-e^{-\frac{ e\psi(z)}{k_BT}} \right) \label{bandbendingpoissoneq}
\end{align}
and using dimensionless variables
 \begin{align}
 \eta =\frac{e \psi}{k_B T} \quad \text{and} \quad \xi=\frac{z-z_s}{L_D} 
\end{align} 
with \(L_D=\sqrt{\epsilon k_BT/4\pi e^2 n_b}\)  we obtain
 \begin{align}
\eta^{\prime \prime}= e^\eta -e^{-\eta}  \text{ .}  \label{diffeqetabb}
 \end{align}
This equation can be integrated once, which gives
 \begin{align}
(\eta^{\prime})^2 =4 \cosh(\eta) +C \text{ .} \label{first_integral}
 \end{align}
The boundary conditions in the bulk \(\eta=0\) and \(\eta^\prime =0\) for \(\xi\rightarrow -\infty\) imply \(C=-4\) so that Eq. (\ref{first_integral}) becomes 
 \begin{align}
\eta^\prime = \sqrt{8} \sinh \left(\frac{\eta}{2}\right) \text{ .} \label{etaprime}
 \end{align}
Integration with the boundary condition at \(z_s\), that is, at \(\xi=0\), \(\eta(0)=\eta_s\) and 
requiring \(\eta \rightarrow 0\) for \(\xi \rightarrow -\infty\) gives
\begin{align}
\eta^\pm(\xi)=\mp 2\ln \left[\pm \tanh \left( \frac{\mp \xi}{\sqrt{2} }+\frac{c^\pm}{2} \right) \right] 
\end{align}
 with 
 \begin{align}
 c^\pm =\pm 2 \text{artanh}\left[ \exp \left(\frac{\mp \eta_s}{2} \right) \right] \text{ ,}
 \end{align}
 where the upper sign is for \(\eta_s>0\) and the lower sign for \(\eta_s<0\).
 
In analogy to what we have done at the boundary of the ESL with the plasma sheath at \(z=z_0\) we 
relate the potential \(\eta_s\) to the total electron surface density in the space charge region. 
From Poisson's equation we obtain for the total electron surface density in the SCR 
 \begin{align}
N^{SCR}=\int_{-\infty}^{z_s} \mathrm{d}z (n-p)=L_D^2 n_b \left. \frac{\mathrm{d}\eta}{\mathrm{d}z} \right|_{-\infty}^{z_s}=L_D n_b \eta^\prime(0) \text{ ,}
 \end{align}
where \( \eta^\prime \) is given by Eq. (\ref{etaprime}), so that
\begin{align}
N^{SCR}=\sqrt{8} L_D n_b \sinh\left(\frac{\eta_s}{2}\right)~, \label{NSC}
\end{align} 
or, 
 \begin{align}
 \eta_s=2 \text{arsinh} \left( \frac{N^{SCR}}{\sqrt{8}L_D n_b}\right) \text{ .} \label{et0eq}
 \end{align}
For a negative space charge \(\eta_s>0\), so that the potential is given by \(\psi(z)=(k_BT/e)\eta^+((z-z_s)/L_D)\) 
and the electron and hole densities are given by  \(n(z)=n_be^{\eta^+((z-z_s)/L_D)}\) and 
\(p(z)=n_b e^{-\eta^+((z-z_s)/L_D)}\), respectively. The relation between \(\psi\) and \(\phi\) is given 
by \(\phi_{bulk}=\phi(z)-\psi(z)\). Since \(\psi(z_s)=(kT/e)\eta_s\) we obtain \(\phi_{bulk}=\phi(z_s)-(kT/e)\eta_s\).
  
Now, quite generally, the excess electrons in the ESL are distributed over the ISR and SCR according to
 \begin{align}
 N=N^{ISR}(\mu)+N^{SCR}(\mu) \label{Ndistrib}~,
 \end{align}
where \(N^{ISR}\) is the surface density of electrons in the ISR, \(\mu\) is the chemical potential in both 
regions, and \(N\) is the total surface density of missing sheath electrons given by Eq.~(\ref{Neq}). The 
total surface density in the ISR is given by \(N^{ISR}=\int_{z_s}^{z_0}\mathrm{d}z n(z)\), where \(n(z)\) is 
calculated with the density functional approach for the graded interface. Requiring continuity of the electron 
density at \(z_s\),
\begin{equation}
\frac{1}{\sqrt{2}}\left(\frac{m_C^\ast k_BT}{\pi \hbar^2} \right)^\frac{3}{2}e^{\frac{1}{k_BT}\left(\mu+e\phi(z_s) \right)}
=n_be^{\eta_s} ~,\label{ncontatzs}
\end{equation}
gives \(\eta_s\) as a function of \(\mu\). From \(\eta_s\) we finally obtain using Eq. (\ref{NSC}) 
\(N^{SCR}(\mu)\). 

For the calculation of the electron distribution and the potential in the refined ESL we use the iteration cycle 
described in the last section with one modification. In step (ii) we solve Eq.~(\ref{Ndistrib}) instead of 
Eq.~(\ref{Nconstraint}) to fix \(\mu\). From \(\mu\) we obtain using Eq. (\ref{ncontatzs}) \(\eta_s\) which 
in turn determines the electron distribution and the potential in the SCR. This gives for each iteration step a 
continuous potential and electron distribution at \(z_s\). As before, the steps (ii) -(iv) are iterated until 
\(\mu\) converges.

At the end of this section let us finally mention two simplifications of the refined ESL model which could be used, 
respectively, for large band gap dielectrics irrespective of the electron affinity and dielectrics 
with small band gap and positive electron affinity. In the former case the intrinsic carrier concentration 
\(n_b\) is very low and the merger with the bulk occurs very deep in the dielectric. Almost all surplus electrons are 
however much closer to the surface where the holes can be neglected. This can be seen from the differential equation 
for \(\eta\). For small \(n_b\) Eq.~(\ref{et0eq}) gives a large \(\eta_s\). As \(\eta\) satisfies a highly nonlinear 
differential equation (\ref{diffeqetabb}), a large \(\eta_s\) implies a steeper gradient of \(\eta\) near the surface 
so that almost all electrons are concentrated close to the surface where neglecting the holes has little effect. Hence, 
for large band gap dielectrics surplus electrons can be filled into a sufficiently large ISR for which the crude ESL 
model of the previous section will be sufficient provided the cut-off \(z_s\) is large enough. The merger with the 
bulk is of course not correctly captured by such an approach.

For dielectrics or semiconductors with small energy gaps and positive electron affinity,
 on the other hand, almost all surplus electrons are deep 
inside the material. It is thus a good approximation to neglect the ISR and to fill all electrons in a SCR. Neglecting 
the surface potential has little effect in this case and using the SCR already for \(z\le0\) gives a good description 
of the electron distribution inside the ESL. The electron density and potential at the surface and in front 
of it can of course not be captured by such an approach. As before \(\psi(z)=\phi(z)-\phi_{bulk}\) with 
\(\phi_{bulk}=\phi(z_s)-(k_BT/e)\eta_s\) where \(\phi(z_s)\) is now the limit of the long range potential 
just inside the dielectric given by \(\phi(z_s)=\phi_w+\chi/e\).
\begin{table}
\caption{Material parameters for the dielectrics considered in this work: dielectric constant \(\epsilon_s\), electron affinity \(\chi\), conduction band effective mass \(m_C^\ast\), valence band effective mass \(m_V^\ast\), band gap \(E_g\). }
\center
\begin{tabular}{l l l l l l}
 &   \(\epsilon_s\) &  \(\chi \) [eV] & \(m_C^\ast\) [\(m_e\)]  &  \(m_V^\ast\) [\(m_e\)] &  \(E_g\) [eV] \\
\hline
MgO  & \(9.8\) [\onlinecite{WFA79}] & \(-0.4\) [\onlinecite{RWK03}] & \(0.4\) [\onlinecite{BZS01}] & & \\
\(\text{Al}_2\text{O}_3\) &  \(9.9\) [\onlinecite{STH00}] & \(2.5\) [\onlinecite{BRB08}] & \(0.4\) [\onlinecite{PSG07}] & \(4.0\)  [\onlinecite{PSG07}] & \(8.8\) [\onlinecite{French90}] \\
\(\text{SiO}_2\)  &  \(3.78\) [\onlinecite{GJ97}] & \(1.3\) [\onlinecite{BRB08}] & \(0.5\) [\onlinecite{LWS98}] & \(0.58\) [\onlinecite{Chanana11}] & \(9.2\) [\onlinecite{BRB08}]  \\
GaAs  &  \(13.1\) [\onlinecite{Zeghbroeck10}] & \(4.07\) [\onlinecite{Zeghbroeck10}] & \(0.067\) [\onlinecite{Zeghbroeck10}] & \(0.45\) [\onlinecite{Zeghbroeck10}] & \(1.42\) [\onlinecite{Zeghbroeck10}] \\
\hline
\end{tabular}
\label{materialtable}
\end{table}

\section{Results}
\label{Results}

We now use the ESL model to calculate for a helium discharge in contact with a MgO, Al$_2$O$_3$, and SiO$_2$ surface 
the potential and the density of excess electrons across the plasma wall. Our main focus lies in the identification
of generic types of electron distributions in the ESL depending on plasma and surface parameters. Unless 
otherwise stated, we use for the plasma density \(n_0=10^{7} cm^{-3}\) and for the electron temperature \(k_BT_e=2eV\). 
The parameters of the dielectric surfaces are given in Table \ref{materialtable}. Preferentially we used
experimental data for the various quantities only if not available we employed theoretical 
values.~\cite{WFA79,RWK03,BZS01,STH00,BRB08,PSG07,French90,GJ97,LWS98,Chanana11,Zeghbroeck10} 

\begin{figure}
\includegraphics[width=\linewidth]{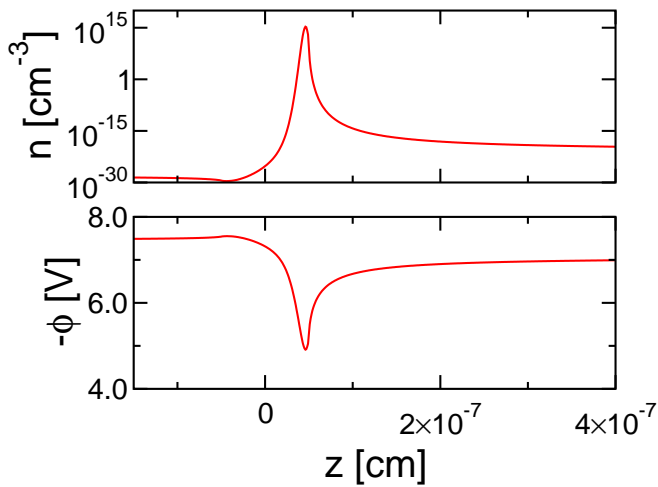}
\caption{Plasma-supplied excess electron density \(n\) (upper panel) and the potential \(-\phi\) (lower panel) it
gives rise to for a MgO surface in contact with a 
helium discharge with \(n_0=10^7 cm^{-3}\) and \(k_BT_e=2eV\) calculated without accounting for a SCR in the
dielectric (crude ESL model). The cut-off of the interface region is \(z_s=-z_0\). As can be seen, almost all of 
the plasma-induced wall charge is located in the well of the image potential in front of the surface.}
\label{figure2}
\end{figure}

First we give typical values for \(z_0\), the position where the ESL merges with the plasma sheath. It is calculated 
from Eq.~(\ref{z0condition}) and should thus depend not only on plasma but also on surface parameters. 
Our results for MgO (\(z_0=6.08\times 10^{-5} cm\)), Al$_3$O$_3$ (\(z_0=6.09\times 10^{-5} cm\)), and SiO$_2$  
(\(z_0=5.14\times 10^{-5} cm\)) indicate however that \(z_0\) is relatively insensitive to \(\epsilon\) which
is the only surface parameter affecting \(z_0\) when the sheath is assumed to be collisionless. Even the 
significantly smaller \(\epsilon\) of SiO$_2$ does not alter \(z_0\) considerably. For the helium discharge considered
\(z_0\) is irrespective of the dielectric always on the order of a micron.

Of particular importance for the distribution of the excess electrons in the ESL is the electron affinity \(\chi\), 
characterizing the offset of the conduction band to the potential just outside. For 
\(\chi <0\) (MgO) the conduction band minimum lies above the potential just outside. It is thus energetically favorable 
for electrons to be located in the image potential in front of the surface. Figure \ref{figure2} showing the electron 
density and the potential in the ESL of MgO verifies this. The energy of an electron in the image potential \(-e\phi\) 
indeed reaches a minimum just in front of the surface at the beginning of the graded interface. For negative electron 
affinity, the excess electrons coming from the plasma thus form an external surface charge in the image potential in 
front of the crystallographic interface. The band bending associated with it is negligible. The external surface charge is very narrow, it can thus be considered as a quasi-two-dimensional electron gas, similar to the surface plasma anticipated by Emeleus and Coulter.\cite{EC87,EC88}
 
\begin{figure}
\includegraphics[width=\linewidth]{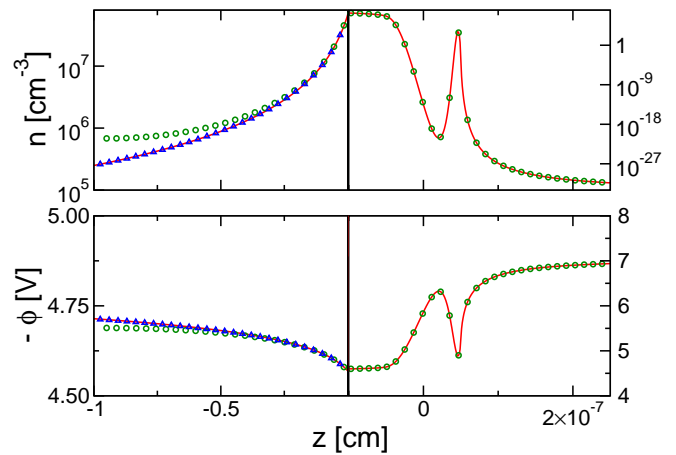}
\caption{Plasma-supplied excess electron density \(n\) (upper panel) and the potential \(-\phi\) (lower panel) 
it gives rise to for an Al$_2$O$_3$ surface 
in contact with a helium discharge with \(n_0=10^7 cm^{-3}\) and \(k_BT_e=2eV\). The red lines show data 
obtained from the refined ESL model accounting for an ISR and a SCR, the boundary between the two was 
put at \(z_s=-3z_0\), the green circles show data for a model which consists only of an ISR with 
cut-off \(z_s=-0.9~cm\) (crude ESL model), and the blue triangles show data for a model consisting
only of a SCR for \(z<0\). Irrespective of the approximation, the plasma-induced wall charge extends deep 
into the bulk. Note the different scales of the axes for the left and right panels. On the spatial scale of 
the SCR shown in the left panels the ISR of the right panels becomes a vertical line.}
\label{figure3}
\end{figure}

For \(\chi > 0\), on the other hand, the conduction band minimum is below the potential just outside.
It is thus energetically favorable for electrons to accumulate inside the dielectric. This can be 
seen in Fig.~\ref{figure3} which shows the electron density and the potential in the refined (red line) 
and simplified ESL (open green circles and blue triangles) for an Al$_2$O$_3$ surface. 
The surface potential consists of an attractive well in front of the surface but the minimum
potential energy for electrons \(-e\phi\) is reached inside the dielectric. Excess electrons coming 
from the plasma are thus mostly located inside the wall and the electron distribution extends deep 
into the bulk. This extended negative space charge also leads to a band bending near the surface. 
Note the different scales of the axes for the left and right panels of Fig. \ref{figure3}. 
On the scale where variations in the SCR are noticeable the ISR is basically a vertical line.

If one neglects the SCR and fills all excess electrons into the ISR (crude ESL) the potential and the electron 
distribution are correctly described at and close to the surface but not far inside the dielectric (open green 
circles) because the ad-hoc cut-off \(z_s\) of the crude ESL leads to an unphysical pile-up of electrons near 
\(z_s\). Hence, only if \(z_s\) is large enough does the crude ESL model give reliable results for the electron 
density and potential in the vicinity of the surface. Filling all electrons in the SCR, on the other hand, cannot
describe the immediate vicinity of the surface correctly which is however on the scale of the SCR an
infinitesimally narrow region. It gives only for \(z<-z_0\) a good description, that 
is, in the region where for \(\chi>0\) indeed most of the electrons are located (blue triangles). 

\begin{figure}
\includegraphics[width=\linewidth]{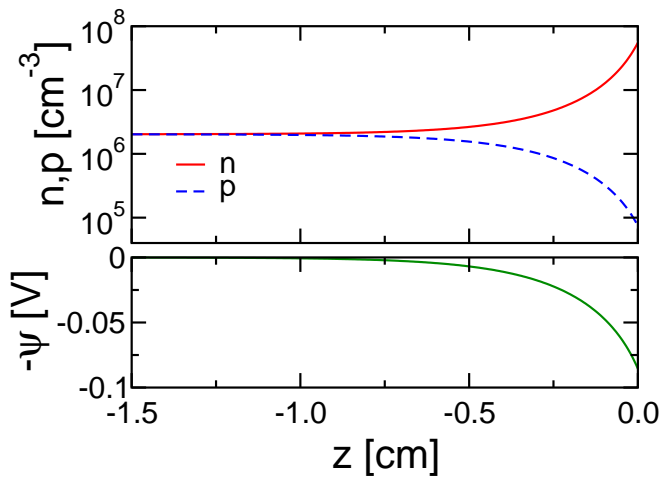}
\caption{Electron density \(n\) and hole density \(p\) at a GaAs surface in contact with a helium discharge with 
\(n_0=10^7 cm^{-3}\) and \(k_BT_e=2eV\) calculated with the refined ESL model without ISR. The plasma-induced
wall charge sits inside the GaAs wall. Deep inside the bulk charge neutrality is achieved by an equal density of
electrons and holes.}
\label{figure3a}
\end{figure}

While the crude ESL model containing only an ISR gives the correct electron density near the surface provided
\(z_s\) is large enough, the merger of the ESL with the bulk can only be described with the refined ESL model 
including the SCR. This is particularly relevant for materials with smaller band gaps and larger intrinsic 
carrier concentrations than MgO, Al$_2$O$_3$, and SiO$_2$. To exemplify this we show in Fig.~\ref{figure3a}
the electron and hole densities (upper panel) as well as the potential \(-\psi\) (lower panel) for a GaAs 
plasma wall, calculated for an ESL containing only a SCR. At the surface the electron density is about 
three orders in magnitude larger than the hole concentration.
Thus, the gas phase plasma offers the possibility to manipulate the electron-hole plasmas by controlling the charge
carrier density - tantamount to doping - in the near surface region of a semiconductor.
 Deep inside the material electron and hole 
concentrations are equal, leading to charge neutrality and a constant potential. The band bending due to the
extended space charge in the ESL is about \(0.09\text{ } eV\). 

Our results for the electron and hole densities and the potential in the dielectric depend of course on the model 
for the SCR. We have used for simplicity the model of an intrinsic semiconductor which is appropriate 
for an undoped semiconductor without impurities. Depending on doping or impurities a variety of models\cite{Lueth92} 
could be used to take material specific aspects into account. In our exploratory calculation we obtain a rather wide 
SCR. Including the effect of impurities, acting as trapping sites in the band gap, would probably reduce 
the depth of the SCR considerably.

To summarize our results up to this point, we find that for negative electron affinity the plasma-induced 
electronic surface charge is located in front of the surface forming a quasi-two-dimensional electron gas,
while for positive electron affinity the surplus electrons form a space-charge 
layer in the dielectric leading to a small bending of the energy bands.

\begin{figure}
\includegraphics[width=\linewidth]{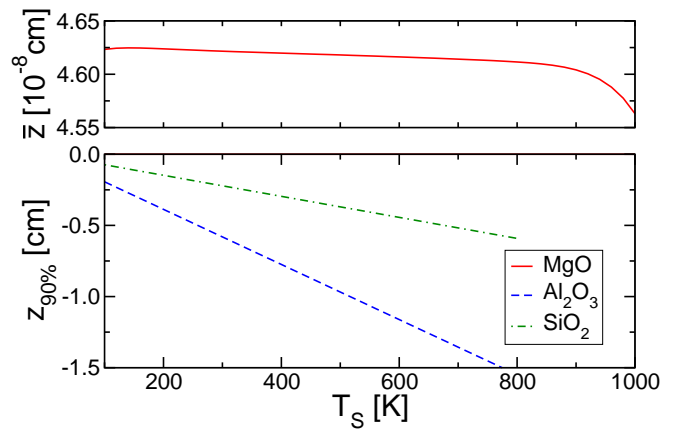}
\caption{Center of gravity \(\bar{z}\) of the plasma-supplied excess electron distribution at a MgO surface (upper 
panel) and the \(z_{90\%}\) value for the electron distribution at an Al$_2$O$_3$ and a SiO$_2$ surface (lower panel),
all in contact with a helium discharge with \(n_0=10^7 cm^{-3}\) and \(k_BT_e=2eV\), as a function of the surface 
temperature \(T_S\). The data shown in the upper and lower panel were obtained, respectively, from the crude 
ESL model and the refined ESL model without an ISR.}
\label{figure5}
\end{figure}

The two distinct types of charge distributions in the ESL are also reflected in the dependence of the 
width of the plasma-supplied electron distribution on the surface temperature. Figure \ref{figure5} shows the center of 
gravity \(\bar{z}\) of the electron distribution for the MgO surface (\(\chi<0\)) and the \(z_{90\%}\) value 
for the surfaces of Al$_2$O$_3$ and SiO$_2$ (\(\chi>0\)), where \(z_{90\%}\) is implicitly defined by
\begin{align}
\int_{z_{90\%}}^0\mathrm{d}z (n(z)-p(z))=0.9N \text{ .}
\end{align}
We use the \(z_{90\%}\) value because it captures the depth of the SCR better than \(\bar{z}\) which depends too 
strongly on the few electrons that penetrate very deep into the bulk.

For negative electron affinity (MgO, shown in the upper panel of Fig.~\ref{figure5}) the external surface charge 
is strongly trapped in the deep image potential so that \(\bar{z}\) changes very little with surface temperature. 
The width of the internal surface charge for dielectrics with positive electron affinity (Al$_2$O$_3$ and SiO$_2$, 
lower panel), however, increases dramatically with surface temperature. This can be understood as follows. 
The restoring force from the positive ions in the sheath binds internal surface charges only weakly to the surface. 
With increasing surface temperature, however, high-lying states in the conduction band get more and more populated. 
Hence, some electrons have rather high kinetic energies, are thus less confined near the surface by the weak 
restoring force, and penetrate therefore deeper into the bulk. As a result, the \(z_{90\%}\) 
value decreases strongly with surface temperature. 

\begin{table}[b]
\caption{Surface density of electrons in the ESL \(N\), wall potential \(\phi_W\), 
plasma sheath - ESL boundary \(z_0\) and the \(z_{90\%}\) value for SiO\(_2\) in contact with a helium discharge 
with \(k_BT_e=2eV\) for different values of the plasma density \(n_0\) .   }
\center
\begin{tabular}{l l l l l}
 \(n_0\) \([10^6 cm^{-3}]\)& \(N\) \([10^6 cm^{-2}]\) &  \(\phi_W\) \([V]\) & \(z_0\) \([10^{-5}cm]\) & \(z_{90\%}\) \([cm]\) \\
\hline 
%5 & 3.10 & -7.07 & 6.11 & -0.099 \\
10 & 4.38 & -7.07 & 5.14 & -0.222 \\
20 & 6.20 & -7.07 & 4.32 & -0.157 \\
50 & 9.80 & -7.07 & 3.44 & -0.099 \\
100 & 13.9 & -7.07 & 2.89 & -0.070 \\
\hline 
\end{tabular}
\label{n0table}
\end{table}

\begin{table}[b]
\caption{Surface density of electrons in the ESL \(N\), wall potential \(\phi_W\), plasma sheath - ESL boundary \(z_0\) and the \(z_{90\%}\) value for SiO\(_2\) in contact with a helium discharge at \(n_0=10^7cm^{-3}\) for different values of the electron temperature \(k_BT_e\).   }
\center
\begin{tabular}{l l l l l}
 \(k_B T_e\) \([eV]\)& \(N\) \([10^6 cm^{-2}]\) &  \(\phi_W\) \([V]\) & \(z_0\) \([10^{-5}cm]\) & \(z_{90\%}\) \([cm]\) \\
\hline 
0.5 & 2.19 & -1.77& 7.27 & -0.444 \\
1 & 3.10 & -3.53 & 6.11 & -0.314 \\
2 & 4.38 & -7.07 & 5.14 & -0.222 \\
5 & 6.93 & -17.7 & 4.086 & -0.140 \\
\hline 
\end{tabular}
\label{kttable}
\end{table}

Let us now turn to the discussion of the influence of the electron temperature \(k_B T_e\) and the plasma density 
\(n_0\) on the properties of the ESL. These two parameters enter through the total surface density of electrons 
\(N\) depleting the sheath and accumulating in the ESL. How \(k_B T_e\) and \(n_0\) affect the interface 
depends therefore on the sheath model and the model used for the interaction between plasma particles and 
the surface. For simplicity we have used a collisionless sheath model and assumed the surface to be a perfect absorber 
for plasma electrons and ions. The results for the properties of the ESL as a function of the plasma parameters are 
thus to be taken only indicative. 
%A complete self-consistent kinetic modeling of the 
%plasma-wall interaction would be required to gain more insights but is beyond the scope of this work.
\begin{figure}
\includegraphics[width=\linewidth]{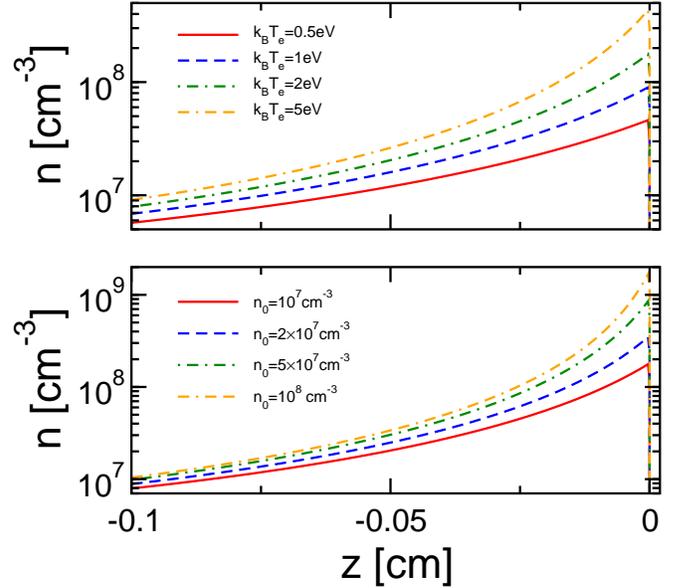}
\caption{Plasma-supplied surplus electron density \(n\) at a SiO$_2$ surface in contact with a helium discharge 
as a function of the 
electron temperature (upper panel, \(n_0=10^7 cm^{-3}\)) and the plasma density (lower panel, 
\(k_BT_e=2eV\)) for \(T_S=300K\). The refined ESL model without an ISR was employed to produce the data.}
\label{figure4}
\end{figure}

The effect of a variation of \(n_0\) and \(k_BT_e\) is most significant for surfaces with positive electron affinity. 
Table \ref{n0table} shows the effect of the plasma density \(n_0\) for a SiO$_2$ surface. If \(n_0\) increases, 
the boundary \(z_0\) between sheath and ESL moves closer to the surface. This, however, does not affect 
the charge distribution much as most of the electrons occupy the SCR inside the dielectric (as shown in 
Fig.~\ref{figure3} for Al$_2$O$_3$). More important is that an increase in \(n_0\) leads to an increase of the 
total surface electron density \(N\). This entails a stronger restoring force from the plasma sheath so that the 
potential well confining the space charge inside the dielectric becomes steeper and the electrons in the SCR of the 
ESL are shifted towards the surface, in other words, the \(z_{90\%}\) value increases with \(n_0\). Mathematically, 
the steeple-like shape of the electron distribution arises because a larger \(N\) leads through Eq. (\ref{et0eq}) 
to a larger \(\eta_s\) which makes the potential steeper at the surface so that the electron distribution is more 
peaked there. This trend can be seen in the lower panel for Fig.~\ref{figure4}.
\begin{figure}
\includegraphics[width=\linewidth]{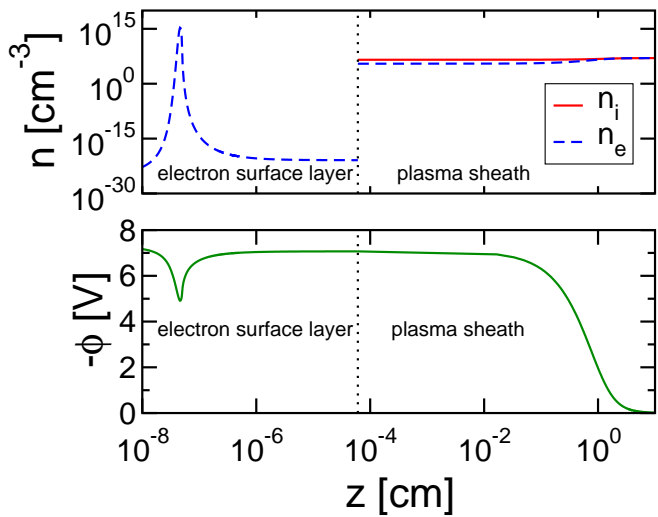}
\caption{Density of plasma-supplied surplus electrons trapped in the ESL as well as electron and ion density in 
the plasma sheath 
(upper panel) and potential (lower panel) for a MgO surface in contact with a helium discharge (\(n_0=10^7 cm^{-3}\) 
and \(k_BT_e=2eV\)). The data were obtained from the crude ESL model.}
\label{figure6}
\end{figure}

A variation of the electron temperature \(k_B T_e\) has similar effects as the variation of the plasma density. 
If \(k_BT_e\) increases, the total surface density of electrons increases also, as can be seen from Table \ref{kttable}. 
As shown in the upper panel of Fig.~\ref{figure4}, this leads again to a steeple-like electron distribution which is
the more concentrated at the surface the higher the electron temperature is.

For a surface with negative electron affinity (MgO) the surplus electrons are strongly bound in the image potential. 
While a variation of \(k_BT_e\) or \(n_0\) changes the total number of surplus electrons per unit area in the same 
way as for a surface with positive electron affinity, the distribution of the surplus electrons within the ESL is not 
affected significantly because of the strong image interaction.

So far we have shown the potential and the electron distribution in the ESL. Now, we will compare potential and charge 
distribution in the ESL with the ones in the plasma sheath. The electron distribution at the interface is the 
quasi-stationary electron gas on top of the charge re-distribution due to the truncation of the solid that guarantees 
flux equality at the sheath-ESL boundary \(z_0\). As already mentioned
not included in this simple model is the flux of plasma electrons and ions in the ESL before the electrons are trapped 
at the surface and the ions recombine with the negative wall charge. The electron and ion densities in this model are
thus discontinuous at \(z_0\). The potential, however, which has been obtained from the integration of Poisson's equation 
is continuous and differentiable everywhere. Between the well of the image potential and \(z_0\) the electron and ion 
flux from the sheath would be important. The neglect of the charge densities associated with these fluxes does however 
not affect the potential because they are too small to have a significant effect.
% over such a short distance. 

\begin{figure}
\includegraphics[width=\linewidth]{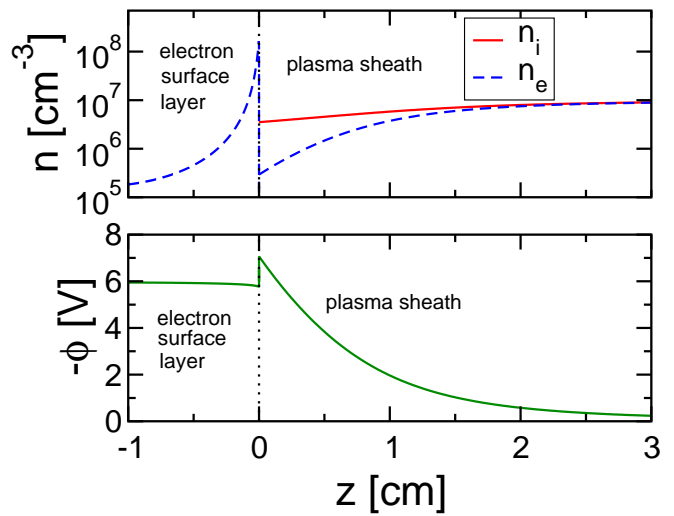}
\caption{Density of plasma-supplied excess electrons in the ESL as well as electron and ion density in the 
plasma sheath (upper panel) 
and potential (lower panel) for a SiO$_2$ surface in contact with a helium discharge (\(n_0=10^7 cm^{-3}\) and 
\(k_BT_e=2eV\)). The refined ESL model without an ISR was employed to produce the data.}
\label{figure6a}
\end{figure}

Figure~\ref{figure6} shows the ESL and the plasma sheath in front of a MgO surface. Due to the negative electron 
affinity, the plasma-supplied surface electrons are bound by the image potential in front of the surface. In 
Fig. \ref{figure6}, 
we plot the electron and ion density (upper panel), as well as the electric potential (lower panel) over the 
distance from the surface \(z\). Far from the surface, the potential approaches the bulk plasma value chosen to 
be zero. In the sheath the potential develops a Coulomb barrier and reaches the wall potential \(\phi_w\) at \(z_0\), 
the distance where the sheath merges with the surface layer (vertical dotted line). The wall potential is the 
potential just outside to which the energies of the bulk states are 
referenced. Closer to the surface the potential follows the attractive image potential while at the surface the 
repulsive potential due to the negative electron affinity prevents the electron from entering the dielectric
(only scarcely seen on the scale of the figure).

In Fig.~\ref{figure6a} we finally plot the electron and ion densities (upper panel) as well as the electric 
potential (lower panel) for SiO$_2$. Note the linear z-axis in contrast to the logarithmic z-axis of 
Fig.~\ref{figure6}. Due to the positive electron affinity, the excess electrons constituting the wall charge 
penetrate deep into the dielectric and occupy therefore the SCR of the ESL. Compared to the variation of the 
electric potential in the sheath the band bending in the dielectric induced by the wall charge is rather 
small as indicated by the variation of \(\phi\) inside the dielectric. This is because \(\epsilon\) is large 
and the width of the SCR is narrow on the scale of the sheath. Only on the scale of the ISR (a vertical 
line at \(z=0\)) the SCR of the ESL is wide. 

\section{Conclusions}
\label{Summary}

We have studied the potential and the charge distribution across the interface of a plasma and a dielectric wall 
treating the plasma-induced wall charge as a quasi-stationary electron gas trapped by and in thermal equilibrium 
with the dielectric. Our approach is based on a model for a graded surface including the offset between 
the potential just outside the dielectric and the conduction band minimum arising from the re-distribution of 
charge due to the truncation of the solid as well as the image potential due to the dielectric mismatch at the 
boundary. The missing electrons from the sheath populate the interface potential and thereby form an electron 
surface layer (ESL) which minimizes the grand canonical potential of wall-thermalized excess electrons and satisfies 
Poisson's equation.

Within this model the boundary between the plasma sheath and the ESL is given by the distance from the crystallographic
interface where the potential for the excess electrons turns from the repulsive sheath potential into the attractive 
surface potential. This distance is typically on the order of a micron. It gives the position of an effective 
wall for plasma electrons and ions and thus the portion of the ESL which lays in front of the surface. Most of the 
surplus electrons trapped in the ESL, that is, the plasma-induced wall charge, will be, however, much closer to the 
surface or even inside the dielectric depending on the electron affinity.

We presented numerical results for the potential and the distribution of the plasma-supplied surplus electrons at 
the interface between a helium discharge and the surfaces of MgO, Al$_2$O$_3$, and SiO$_2$, respectively. 
The electron distribution within the ESL strongly depends on the electron affinity. For 
negative electron affinity, the conduction band minimum is above the potential just outside the dielectric. Hence, 
it is energetically unfavorable for electrons to penetrate into the bulk and the surface electrons are bound in the 
image potential in front of the surface. In this case, their spatial profile changes little over a variation of the 
surface temperature or the plasma parameters. For positive electron affinity the conduction band minimum is below 
the potential just outside the dielectric and the surface-bound electrons accumulate inside the wall. The space 
charge in the bulk broadens if the surface temperature is increased and becomes more peaked if the total surface 
density of the electrons missing in the sheath is raised through either an increase in the plasma density or the 
electron temperature. 
%For plasma walls with positive electron affinity the plasma induces also an additional band 
%bending. 

Separating the ESL into an interface specific and a space charge region and modeling the bulk of the dielectric 
as an intrinsic semiconductor we also investigated how the ESL merges with the bulk of the dielectric. This 
is particularly important for dielectrics with small energy gaps and positive electron affinities where excess 
electrons coming from the sheath accumulate not in the image potential in front of the surface but deep inside 
the wall. In this case the wall charge may also induce a significant band bending. 

Whereas the crude ESL model 
we proposed neglects the space charge deep inside the bulk of the wall and is thus only applicable to large 
band gap dielectrics with negative electron affinity where basically the whole plasma-induced wall charge is 
trapped in the
image potential in front of the surface, the refined ESL model keeping the interface specific as well as the 
space charge region of the ESL provides a quantitative description of the whole spatial structure of the extended 
charge double-layer which forms at a dielectric plasma wall as a result of the electrons in the ESL and the
positive space charge in the plasma sheath. 

The ESL can be regarded as that part of the plasma sheath which is inside the plasma wall. It is thus the 
ultimate boundary of a bounded gas discharge and constitutes, depending on the electron affinity either 
a quasi two-dimensional electron plasma in front of the wall or an electron(-hole) plasma inside the wall.

{\it Acknowledgments.}
This work was supported by the Deutsche Forschungsgemeinschaft (DFG) through the projects B10 and A5 of the 
transregional collaborative research center SFB/TRR 24. 

%\bibliography{./ref} 

\begin{thebibliography}{60}
\expandafter\ifx\csname natexlab\endcsname\relax\def\natexlab#1{#1}\fi
\expandafter\ifx\csname bibnamefont\endcsname\relax
  \def\bibnamefont#1{#1}\fi
\expandafter\ifx\csname bibfnamefont\endcsname\relax
  \def\bibfnamefont#1{#1}\fi
\expandafter\ifx\csname citenamefont\endcsname\relax
  \def\citenamefont#1{#1}\fi
\expandafter\ifx\csname url\endcsname\relax
  \def\url#1{\texttt{#1}}\fi
\expandafter\ifx\csname urlprefix\endcsname\relax\def\urlprefix{URL }\fi
\providecommand{\bibinfo}[2]{#2}
\providecommand{\eprint}[2][]{\url{#2}}

\bibitem[{\citenamefont{Ishihara}(2007)}]{Ishihara07}
\bibinfo{author}{\bibfnamefont{O.}~\bibnamefont{Ishihara}},
  \bibinfo{journal}{J. Phys. D: Appl. Phys} \textbf{\bibinfo{volume}{40}},
  \bibinfo{pages}{R121} (\bibinfo{year}{2007}).

\bibitem[{\citenamefont{Fortov et~al.}(2005)\citenamefont{Fortov, Ivlev,
  Khrapak, Khrapak, and Morfill}}]{FIK05}
\bibinfo{author}{\bibfnamefont{V.~E.} \bibnamefont{Fortov}},
  \bibinfo{author}{\bibfnamefont{A.~V.} \bibnamefont{Ivlev}},
  \bibinfo{author}{\bibfnamefont{S.~A.} \bibnamefont{Khrapak}},
  \bibinfo{author}{\bibfnamefont{A.~G.} \bibnamefont{Khrapak}},
  \bibnamefont{and} \bibinfo{author}{\bibfnamefont{G.~E.}
  \bibnamefont{Morfill}}, \bibinfo{journal}{Physics Reports}
  \textbf{\bibinfo{volume}{421}}, \bibinfo{pages}{1} (\bibinfo{year}{2005}).

\bibitem[{\citenamefont{Mendis}(2002)}]{Mendis02}
\bibinfo{author}{\bibfnamefont{D.~A.} \bibnamefont{Mendis}},
  \bibinfo{journal}{Plasma Sources Sci. Technol.}
  \textbf{\bibinfo{volume}{11}}, \bibinfo{pages}{A219} (\bibinfo{year}{2002}).

\bibitem[{\citenamefont{Kogelschatz}(2003)}]{Kogelschatz03}
\bibinfo{author}{\bibfnamefont{U.}~\bibnamefont{Kogelschatz}},
  \bibinfo{journal}{Plasma Chemistry and Plasma Processing}
  \textbf{\bibinfo{volume}{23}}, \bibinfo{pages}{1} (\bibinfo{year}{2003}).

\bibitem[{\citenamefont{Stollenwerk et~al.}(2006)\citenamefont{Stollenwerk,
  Amiranashvili, Boeuf, and Purwins}}]{SAB06}
\bibinfo{author}{\bibfnamefont{L.}~\bibnamefont{Stollenwerk}},
  \bibinfo{author}{\bibfnamefont{S.}~\bibnamefont{Amiranashvili}},
  \bibinfo{author}{\bibfnamefont{J.-P.} \bibnamefont{Boeuf}}, \bibnamefont{and}
  \bibinfo{author}{\bibfnamefont{H.-G.} \bibnamefont{Purwins}},
  \bibinfo{journal}{Phys. Rev. Lett.} \textbf{\bibinfo{volume}{96}},
  \bibinfo{pages}{255001} (\bibinfo{year}{2006}).

\bibitem[{\citenamefont{Stollenwerk et~al.}(2007)\citenamefont{Stollenwerk,
  Laven, and Purwins}}]{SLP07}
\bibinfo{author}{\bibfnamefont{L.}~\bibnamefont{Stollenwerk}},
  \bibinfo{author}{\bibfnamefont{J.~G.} \bibnamefont{Laven}}, \bibnamefont{and}
  \bibinfo{author}{\bibfnamefont{H.-G.} \bibnamefont{Purwins}},
  \bibinfo{journal}{Phys. Rev. Lett.} \textbf{\bibinfo{volume}{98}},
  \bibinfo{pages}{255001} (\bibinfo{year}{2007}).

\bibitem[{\citenamefont{Dussart et~al.}(2010)\citenamefont{Dussart, Overzet,
  Lefaucheux, Dufour, Kulsreshath, Mandra, Tillocher, Aubry, Dozias, Ranson
  et~al.}}]{DOL10}
\bibinfo{author}{\bibfnamefont{R.}~\bibnamefont{Dussart}},
  \bibinfo{author}{\bibfnamefont{L.}~\bibnamefont{Overzet}},
  \bibinfo{author}{\bibfnamefont{P.}~\bibnamefont{Lefaucheux}},
  \bibinfo{author}{\bibfnamefont{T.}~\bibnamefont{Dufour}},
  \bibinfo{author}{\bibfnamefont{M.}~\bibnamefont{Kulsreshath}},
  \bibinfo{author}{\bibfnamefont{M.}~\bibnamefont{Mandra}},
  \bibinfo{author}{\bibfnamefont{T.}~\bibnamefont{Tillocher}},
  \bibinfo{author}{\bibfnamefont{O.}~\bibnamefont{Aubry}},
  \bibinfo{author}{\bibfnamefont{S.}~\bibnamefont{Dozias}},
  \bibinfo{author}{\bibfnamefont{P.}~\bibnamefont{Ranson}},
  \bibnamefont{et~al.}, \bibinfo{journal}{Eur. Phys. J. D}
  \textbf{\bibinfo{volume}{60}}, \bibinfo{pages}{601} (\bibinfo{year}{2010}).

\bibitem[{\citenamefont{Wagner et~al.}(2010)\citenamefont{Wagner, Tchertchian,
  and Eden}}]{WTE10}
\bibinfo{author}{\bibfnamefont{C.~J.} \bibnamefont{Wagner}},
  \bibinfo{author}{\bibfnamefont{P.~A.} \bibnamefont{Tchertchian}},
  \bibnamefont{and} \bibinfo{author}{\bibfnamefont{J.~G.} \bibnamefont{Eden}},
  \bibinfo{journal}{Appl. Phys. Lett.} \textbf{\bibinfo{volume}{97}},
  \bibinfo{pages}{134102} (\bibinfo{year}{2010}).

\bibitem[{\citenamefont{Becker et~al.}(2006)\citenamefont{Becker, Schoenbach,
  and Eden}}]{BSE06}
\bibinfo{author}{\bibfnamefont{K.~H.} \bibnamefont{Becker}},
  \bibinfo{author}{\bibfnamefont{K.~H.} \bibnamefont{Schoenbach}},
  \bibnamefont{and} \bibinfo{author}{\bibfnamefont{J.~G.} \bibnamefont{Eden}},
  \bibinfo{journal}{J. Phys. D: Appl. Phys} \textbf{\bibinfo{volume}{39}},
  \bibinfo{pages}{R55} (\bibinfo{year}{2006}).

\bibitem[{\citenamefont{Ostrom and Eden}(2005)}]{OE05}
\bibinfo{author}{\bibfnamefont{N.~P.} \bibnamefont{Ostrom}} \bibnamefont{and}
  \bibinfo{author}{\bibfnamefont{J.~G.} \bibnamefont{Eden}},
  \bibinfo{journal}{Appl. Phys. Lett.} \textbf{\bibinfo{volume}{87}},
  \bibinfo{pages}{141101} (\bibinfo{year}{2005}).

\bibitem[{\citenamefont{Kushner}(2005)}]{Kushner05}
\bibinfo{author}{\bibfnamefont{M.~J.} \bibnamefont{Kushner}},
  \bibinfo{journal}{J. Phys. D: Appl. Phys} \textbf{\bibinfo{volume}{38}},
  \bibinfo{pages}{1633} (\bibinfo{year}{2005}).

\bibitem[{\citenamefont{Franklin}(1976)}]{Franklin76}
\bibinfo{author}{\bibfnamefont{R.~N.} \bibnamefont{Franklin}},
  \emph{\bibinfo{title}{Plasma phenomena in gas discharges}}
  (\bibinfo{publisher}{Clarendon Press}, \bibinfo{address}{Oxford},
  \bibinfo{year}{1976}).

\bibitem[{\citenamefont{Lieberman and Lichtenberg}(2005)}]{LL05}
\bibinfo{author}{\bibfnamefont{M.~A.} \bibnamefont{Lieberman}}
  \bibnamefont{and} \bibinfo{author}{\bibfnamefont{A.~J.}
  \bibnamefont{Lichtenberg}}, \emph{\bibinfo{title}{Principles of plasma
  discharges and materials processing}}
  (\bibinfo{publisher}{Wiley-Interscience}, \bibinfo{address}{New York},
  \bibinfo{year}{2005}).

\bibitem[{\citenamefont{Riemann}(1991)}]{Riemann91}
\bibinfo{author}{\bibfnamefont{K.-U.} \bibnamefont{Riemann}},
  \bibinfo{journal}{J. Phys. D: Appl. Phys} \textbf{\bibinfo{volume}{24}},
  \bibinfo{pages}{493} (\bibinfo{year}{1991}).

\bibitem[{\citenamefont{Emeleus and Coulter}(1987)}]{EC87}
\bibinfo{author}{\bibfnamefont{K.~G.} \bibnamefont{Emeleus}} \bibnamefont{and}
  \bibinfo{author}{\bibfnamefont{J.~R.~M.} \bibnamefont{Coulter}},
  \bibinfo{journal}{Int. J. Electronics} \textbf{\bibinfo{volume}{62}},
  \bibinfo{pages}{225} (\bibinfo{year}{1987}).

\bibitem[{\citenamefont{Emeleus and Coulter}(1988)}]{EC88}
\bibinfo{author}{\bibfnamefont{K.~G.} \bibnamefont{Emeleus}} \bibnamefont{and}
  \bibinfo{author}{\bibfnamefont{J.~R.~M.} \bibnamefont{Coulter}},
  \bibinfo{journal}{IEE Proceedings} \textbf{\bibinfo{volume}{135}},
  \bibinfo{pages}{76} (\bibinfo{year}{1988}).

\bibitem[{\citenamefont{Behnke et~al.}(1997)\citenamefont{Behnke, Bindemann,
  Deutsch, and Becker}}]{BBD97}
\bibinfo{author}{\bibfnamefont{J.~F.} \bibnamefont{Behnke}},
  \bibinfo{author}{\bibfnamefont{T.}~\bibnamefont{Bindemann}},
  \bibinfo{author}{\bibfnamefont{H.}~\bibnamefont{Deutsch}}, \bibnamefont{and}
  \bibinfo{author}{\bibfnamefont{K.}~\bibnamefont{Becker}},
  \bibinfo{journal}{Contrib. Plasma Phys.} \textbf{\bibinfo{volume}{37}},
  \bibinfo{pages}{345} (\bibinfo{year}{1997}).

\bibitem[{\citenamefont{Uhrlandt et~al.}(2000)\citenamefont{Uhrlandt, Schmidt,
  Behnke, and Bindemann}}]{USB00}
\bibinfo{author}{\bibfnamefont{D.}~\bibnamefont{Uhrlandt}},
  \bibinfo{author}{\bibfnamefont{M.}~\bibnamefont{Schmidt}},
  \bibinfo{author}{\bibfnamefont{J.~F.} \bibnamefont{Behnke}},
  \bibnamefont{and}
  \bibinfo{author}{\bibfnamefont{T.}~\bibnamefont{Bindemann}},
  \bibinfo{journal}{J. Phys. D: Appl. Phys} \textbf{\bibinfo{volume}{33}},
  \bibinfo{pages}{2475} (\bibinfo{year}{2000}).

\bibitem[{\citenamefont{Golubovskii et~al.}(2002)\citenamefont{Golubovskii,
  Maiorov, Behnke, and Behnke}}]{GMB02}
\bibinfo{author}{\bibfnamefont{Y.~B.} \bibnamefont{Golubovskii}},
  \bibinfo{author}{\bibfnamefont{V.~A.} \bibnamefont{Maiorov}},
  \bibinfo{author}{\bibfnamefont{J.}~\bibnamefont{Behnke}}, \bibnamefont{and}
  \bibinfo{author}{\bibfnamefont{J.~F.} \bibnamefont{Behnke}},
  \bibinfo{journal}{J. Phys. D: Appl. Phys} \textbf{\bibinfo{volume}{35}},
  \bibinfo{pages}{751} (\bibinfo{year}{2002}).

\bibitem[{\citenamefont{Bronold et~al.}(2009)\citenamefont{Bronold, Deutsch,
  and Fehske}}]{BDF09}
\bibinfo{author}{\bibfnamefont{F.~X.} \bibnamefont{Bronold}},
  \bibinfo{author}{\bibfnamefont{H.}~\bibnamefont{Deutsch}}, \bibnamefont{and}
  \bibinfo{author}{\bibfnamefont{H.}~\bibnamefont{Fehske}},
  \bibinfo{journal}{Eur. Phys. J. D} \textbf{\bibinfo{volume}{54}},
  \bibinfo{pages}{519} (\bibinfo{year}{2009}).

\bibitem[{\citenamefont{Heinisch
  et~al.}(2010{\natexlab{a}})\citenamefont{Heinisch, Bronold, and
  Fehske}}]{HBF10a}
\bibinfo{author}{\bibfnamefont{R.~L.} \bibnamefont{Heinisch}},
  \bibinfo{author}{\bibfnamefont{F.~X.} \bibnamefont{Bronold}},
  \bibnamefont{and} \bibinfo{author}{\bibfnamefont{H.}~\bibnamefont{Fehske}},
  \bibinfo{journal}{Phys. Rev. B} \textbf{\bibinfo{volume}{81}},
  \bibinfo{pages}{155420} (\bibinfo{year}{2010}{\natexlab{a}}).

\bibitem[{\citenamefont{Heinisch
  et~al.}(2010{\natexlab{b}})\citenamefont{Heinisch, Bronold, and
  Fehske}}]{HBF10b}
\bibinfo{author}{\bibfnamefont{R.~L.} \bibnamefont{Heinisch}},
  \bibinfo{author}{\bibfnamefont{F.~X.} \bibnamefont{Bronold}},
  \bibnamefont{and} \bibinfo{author}{\bibfnamefont{H.}~\bibnamefont{Fehske}},
  \bibinfo{journal}{Phys. Rev. B} \textbf{\bibinfo{volume}{82}},
  \bibinfo{pages}{125408} (\bibinfo{year}{2010}{\natexlab{b}}).

\bibitem[{\citenamefont{Bronold et~al.}(2011)\citenamefont{Bronold, Heinisch,
  Marbach, and Fehske}}]{BHMF11}
\bibinfo{author}{\bibfnamefont{F.~X.} \bibnamefont{Bronold}},
  \bibinfo{author}{\bibfnamefont{R.~L.} \bibnamefont{Heinisch}},
  \bibinfo{author}{\bibfnamefont{J.}~\bibnamefont{Marbach}}, \bibnamefont{and}
  \bibinfo{author}{\bibfnamefont{H.}~\bibnamefont{Fehske}},
  \bibinfo{journal}{IEEE Transactions on Plasma Science}
  \textbf{\bibinfo{volume}{39}}, \bibinfo{pages}{644} (\bibinfo{year}{2011}).

\bibitem[{\citenamefont{Heinisch et~al.}(2011)\citenamefont{Heinisch, Bronold,
  and Fehske}}]{HBF11}
\bibinfo{author}{\bibfnamefont{R.~L.} \bibnamefont{Heinisch}},
  \bibinfo{author}{\bibfnamefont{F.~X.} \bibnamefont{Bronold}},
  \bibnamefont{and} \bibinfo{author}{\bibfnamefont{H.}~\bibnamefont{Fehske}},
  \bibinfo{journal}{Phys. Rev. B} \textbf{\bibinfo{volume}{83}},
  \bibinfo{pages}{195407} (\bibinfo{year}{2011}).

\bibitem[{\citenamefont{Bronold et~al.}(2008)\citenamefont{Bronold, Fehske,
  Kersten, and Deutsch}}]{BFKD08}
\bibinfo{author}{\bibfnamefont{F.~X.} \bibnamefont{Bronold}},
  \bibinfo{author}{\bibfnamefont{H.}~\bibnamefont{Fehske}},
  \bibinfo{author}{\bibfnamefont{H.}~\bibnamefont{Kersten}}, \bibnamefont{and}
  \bibinfo{author}{\bibfnamefont{H.}~\bibnamefont{Deutsch}},
  \bibinfo{journal}{Phys. Rev. Lett.} \textbf{\bibinfo{volume}{101}},
  \bibinfo{pages}{175002} (\bibinfo{year}{2008}).

\bibitem[{\citenamefont{Cole and Cohen}(1969)}]{CC69}
\bibinfo{author}{\bibfnamefont{M.~W.} \bibnamefont{Cole}} \bibnamefont{and}
  \bibinfo{author}{\bibfnamefont{M.~H.} \bibnamefont{Cohen}},
  \bibinfo{journal}{Phys. Rev. Lett.} \textbf{\bibinfo{volume}{23}},
  \bibinfo{pages}{1238} (\bibinfo{year}{1969}).

\bibitem[{\citenamefont{Baumeier et~al.}(2007)\citenamefont{Baumeier, Kruger,
  and Pollmann}}]{BKP07}
\bibinfo{author}{\bibfnamefont{B.}~\bibnamefont{Baumeier}},
  \bibinfo{author}{\bibfnamefont{P.}~\bibnamefont{Kruger}}, \bibnamefont{and}
  \bibinfo{author}{\bibfnamefont{J.}~\bibnamefont{Pollmann}},
  \bibinfo{journal}{Phys. Rev. B} \textbf{\bibinfo{volume}{76}},
  \bibinfo{pages}{205404} (\bibinfo{year}{2007}).

\bibitem[{\citenamefont{Rohlfing et~al.}(2003)\citenamefont{Rohlfing, Wang,
  Kruger, and Pollmann}}]{RWK03}
\bibinfo{author}{\bibfnamefont{M.}~\bibnamefont{Rohlfing}},
  \bibinfo{author}{\bibfnamefont{N.-P.} \bibnamefont{Wang}},
  \bibinfo{author}{\bibfnamefont{P.}~\bibnamefont{Kruger}}, \bibnamefont{and}
  \bibinfo{author}{\bibfnamefont{J.}~\bibnamefont{Pollmann}},
  \bibinfo{journal}{Phys. Rev. Lett.} \textbf{\bibinfo{volume}{91}},
  \bibinfo{pages}{256802} (\bibinfo{year}{2003}).

\bibitem[{\citenamefont{Silkin et~al.}(2009)\citenamefont{Silkin, Zhao, Guinea,
  Chulkov, Echenique, and Petek}}]{SZG09}
\bibinfo{author}{\bibfnamefont{V.~M.} \bibnamefont{Silkin}},
  \bibinfo{author}{\bibfnamefont{J.}~\bibnamefont{Zhao}},
  \bibinfo{author}{\bibfnamefont{F.}~\bibnamefont{Guinea}},
  \bibinfo{author}{\bibfnamefont{E.~V.} \bibnamefont{Chulkov}},
  \bibinfo{author}{\bibfnamefont{P.~M.} \bibnamefont{Echenique}},
  \bibnamefont{and} \bibinfo{author}{\bibfnamefont{H.}~\bibnamefont{Petek}},
  \bibinfo{journal}{Phys. Rev. B} \textbf{\bibinfo{volume}{80}},
  \bibinfo{pages}{121408(R)} (\bibinfo{year}{2009}).

\bibitem[{\citenamefont{Silkin et~al.}(1999)\citenamefont{Silkin, Chulkov, and
  Echenique}}]{SCE99}
\bibinfo{author}{\bibfnamefont{V.~M.} \bibnamefont{Silkin}},
  \bibinfo{author}{\bibfnamefont{E.~V.} \bibnamefont{Chulkov}},
  \bibnamefont{and} \bibinfo{author}{\bibfnamefont{P.~M.}
  \bibnamefont{Echenique}}, \bibinfo{journal}{Phys. Rev. B}
  \textbf{\bibinfo{volume}{60}}, \bibinfo{pages}{7820} (\bibinfo{year}{1999}).

\bibitem[{\citenamefont{Hoefer et~al.}(1997)\citenamefont{Hoefer, Shumay,
  Reuss, Thomann, Wallauer, and Fauster}}]{HSR97}
\bibinfo{author}{\bibfnamefont{U.}~\bibnamefont{Hoefer}},
  \bibinfo{author}{\bibfnamefont{I.~L.} \bibnamefont{Shumay}},
  \bibinfo{author}{\bibfnamefont{C.}~\bibnamefont{Reuss}},
  \bibinfo{author}{\bibfnamefont{U.}~\bibnamefont{Thomann}},
  \bibinfo{author}{\bibfnamefont{W.}~\bibnamefont{Wallauer}}, \bibnamefont{and}
  \bibinfo{author}{\bibfnamefont{T.}~\bibnamefont{Fauster}},
  \bibinfo{journal}{Science} \textbf{\bibinfo{volume}{277}},
  \bibinfo{pages}{1480} (\bibinfo{year}{1997}).

\bibitem[{\citenamefont{Fauster}(1994)}]{Fauster94}
\bibinfo{author}{\bibfnamefont{T.}~\bibnamefont{Fauster}},
  \bibinfo{journal}{Appl. Phys. A} \textbf{\bibinfo{volume}{59}},
  \bibinfo{pages}{479} (\bibinfo{year}{1994}).

\bibitem[{\citenamefont{Stern}(1978)}]{Stern78}
\bibinfo{author}{\bibfnamefont{F.}~\bibnamefont{Stern}},
  \bibinfo{journal}{Phys. Rev. B} \textbf{\bibinfo{volume}{17}},
  \bibinfo{pages}{5009} (\bibinfo{year}{1978}).

\bibitem[{\citenamefont{Stern and DasSarma}(1984)}]{SS84}
\bibinfo{author}{\bibfnamefont{F.}~\bibnamefont{Stern}} \bibnamefont{and}
  \bibinfo{author}{\bibfnamefont{S.} \bibnamefont{DasSarma}},
  \bibinfo{journal}{Phys. Rev. B} \textbf{\bibinfo{volume}{30}},
  \bibinfo{pages}{840} (\bibinfo{year}{1984}).

\bibitem[{\citenamefont{Planelles and Movilla}(2006)}]{PM06}
\bibinfo{author}{\bibfnamefont{J.}~\bibnamefont{Planelles}} \bibnamefont{and}
  \bibinfo{author}{\bibfnamefont{J.~L.} \bibnamefont{Movilla}},
  \bibinfo{journal}{Phys. Rev. B} \textbf{\bibinfo{volume}{73}},
  \bibinfo{pages}{235350} (\bibinfo{year}{2006}).

\bibitem[{\citenamefont{Tkharev and Danilyuk}(1985)}]{TD85}
\bibinfo{author}{\bibfnamefont{E.~E.} \bibnamefont{Tkharev}} \bibnamefont{and}
  \bibinfo{author}{\bibfnamefont{A.~L.} \bibnamefont{Danilyuk}},
  \bibinfo{journal}{Vacuum} \textbf{\bibinfo{volume}{35}}, \bibinfo{pages}{183}
  (\bibinfo{year}{1985}).

\bibitem[{\citenamefont{Mermin}(1965)}]{Mermin65}
\bibinfo{author}{\bibfnamefont{N.~D.} \bibnamefont{Mermin}},
  \bibinfo{journal}{Phys. Rev.} \textbf{\bibinfo{volume}{137A}},
  \bibinfo{pages}{1441} (\bibinfo{year}{1965}).

\bibitem[{\citenamefont{Kohn and Sham}(1965)}]{KS65}
\bibinfo{author}{\bibfnamefont{W.}~\bibnamefont{Kohn}} \bibnamefont{and}
  \bibinfo{author}{\bibfnamefont{L.~J.} \bibnamefont{Sham}},
  \bibinfo{journal}{Phys. Rev.} \textbf{\bibinfo{volume}{140A}},
  \bibinfo{pages}{1133} (\bibinfo{year}{1965}).

\bibitem[{\citenamefont{Jones and Gunnarson}(1989)}]{JG89}
\bibinfo{author}{\bibfnamefont{R.~O.} \bibnamefont{Jones}} \bibnamefont{and}
  \bibinfo{author}{\bibfnamefont{O.}~\bibnamefont{Gunnarson}},
  \bibinfo{journal}{Rev. Mod. Phys.} \textbf{\bibinfo{volume}{61}},
  \bibinfo{pages}{689} (\bibinfo{year}{1989}).

\bibitem[{\citenamefont{Jennings et~al.}(1999)\citenamefont{Jennings, Jones,
  and Weinert}}]{JJW88}
\bibinfo{author}{\bibfnamefont{P.~J.} \bibnamefont{Jennings}},
  \bibinfo{author}{\bibfnamefont{R.~O.} \bibnamefont{Jones}}, \bibnamefont{and}
  \bibinfo{author}{\bibfnamefont{M.}~\bibnamefont{Weinert}},
  \bibinfo{journal}{Phys. Rev. B} \textbf{\bibinfo{volume}{37}},
  \bibinfo{pages}{6113} (\bibinfo{year}{1988}).

\bibitem[{\citenamefont{Gavrilenko et~al.}(2010)\citenamefont{Gavrilenko,
  McKinney, and Gavrilenko}}]{GMG10}
\bibinfo{author}{\bibfnamefont{A.~V.} \bibnamefont{Gavrilenko}},
  \bibinfo{author}{\bibfnamefont{C.~S.} \bibnamefont{McKinney}},
  \bibnamefont{and} \bibinfo{author}{\bibfnamefont{V.~I.}
  \bibnamefont{Gavrilenko}}, \bibinfo{journal}{Phys. Rev. B}
  \textbf{\bibinfo{volume}{82}}, \bibinfo{pages}{155426}
  (\bibinfo{year}{2010}).

\bibitem[{\citenamefont{L\"{u}th}(1992)}]{Lueth92}
\bibinfo{author}{\bibfnamefont{H.}~\bibnamefont{L\"{u}th}},
  \emph{\bibinfo{title}{Solid Surfaces, Interfaces and Thin Films}}
  (\bibinfo{publisher}{Springer Verlag}, \bibinfo{address}{Berlin},
  \bibinfo{year}{1992}).

\bibitem[{\citenamefont{Tung}(2001)}]{Tung01}
\bibinfo{author}{\bibfnamefont{R.~T.} \bibnamefont{Tung}},
  \bibinfo{journal}{Material Science and Engineering}
  \textbf{\bibinfo{volume}{R35}}, \bibinfo{pages}{1} (\bibinfo{year}{2001}).

\bibitem[{\citenamefont{Cahen and Kahn}(2003)}]{CK03}
\bibinfo{author}{\bibfnamefont{D.}~\bibnamefont{Cahen}} \bibnamefont{and}
  \bibinfo{author}{\bibfnamefont{A.}~\bibnamefont{Kahn}},
  \bibinfo{journal}{Advanced Materials} \textbf{\bibinfo{volume}{15}},
  \bibinfo{pages}{271} (\bibinfo{year}{2003}).

\bibitem[{\citenamefont{Himpsel et~al.}(1979)\citenamefont{Himpsel, Knapp,
  VanVechten, and Eastman}}]{HKV79}
\bibinfo{author}{\bibfnamefont{F.~J.} \bibnamefont{Himpsel}},
  \bibinfo{author}{\bibfnamefont{J.~A.} \bibnamefont{Knapp}},
  \bibinfo{author}{\bibfnamefont{J.~A.} \bibnamefont{VanVechten}},
  \bibnamefont{and} \bibinfo{author}{\bibfnamefont{D.~E.}
  \bibnamefont{Eastman}}, \bibinfo{journal}{Phys. Rev. B}
  \textbf{\bibinfo{volume}{20}}, \bibinfo{pages}{624} (\bibinfo{year}{1979}).

\bibitem[{\citenamefont{Loh et~al.}(1999)\citenamefont{Loh, Sakaguchi, Gamo,
  Tagawa, Sugino, and Ando}}]{LSG99}
\bibinfo{author}{\bibfnamefont{K.~P.} \bibnamefont{Loh}},
  \bibinfo{author}{\bibfnamefont{I.}~\bibnamefont{Sakaguchi}},
  \bibinfo{author}{\bibfnamefont{M.~N.} \bibnamefont{Gamo}},
  \bibinfo{author}{\bibfnamefont{S.}~\bibnamefont{Tagawa}},
  \bibinfo{author}{\bibfnamefont{T.}~\bibnamefont{Sugino}}, \bibnamefont{and}
  \bibinfo{author}{\bibfnamefont{T.}~\bibnamefont{Ando}},
  \bibinfo{journal}{Appl. Phys. Lett.} \textbf{\bibinfo{volume}{74}},
  \bibinfo{pages}{28} (\bibinfo{year}{1999}).

\bibitem[{\citenamefont{Cui et~al.}(1998)\citenamefont{Cui, Ristein, and
  Ley}}]{CRL98}
\bibinfo{author}{\bibfnamefont{J.~B.} \bibnamefont{Cui}},
  \bibinfo{author}{\bibfnamefont{J.}~\bibnamefont{Ristein}}, \bibnamefont{and}
  \bibinfo{author}{\bibfnamefont{L.}~\bibnamefont{Ley}},
  \bibinfo{journal}{Phys. Rev. Lett.} \textbf{\bibinfo{volume}{81}},
  \bibinfo{pages}{429} (\bibinfo{year}{1998}).

\bibitem[{\citenamefont{Maier et~al.}(2001)\citenamefont{Maier, Ristein, and
  Ley}}]{FRL01}
\bibinfo{author}{\bibfnamefont{F.}~\bibnamefont{Maier}},
  \bibinfo{author}{\bibfnamefont{J.}~\bibnamefont{Ristein}}, \bibnamefont{and}
  \bibinfo{author}{\bibfnamefont{L.}~\bibnamefont{Ley}},
  \bibinfo{journal}{Phys. Rev. B} \textbf{\bibinfo{volume}{64}},
  \bibinfo{pages}{165411} (\bibinfo{year}{2001}).

\bibitem[{\citenamefont{Jackson}(1998)}]{Jackson98}
\bibinfo{author}{\bibfnamefont{J.~D.} \bibnamefont{Jackson}},
  \emph{\bibinfo{title}{Classical electrodynamics}}
  (\bibinfo{publisher}{Wiley}, \bibinfo{address}{New York},
  \bibinfo{year}{1998}).

\bibitem[{\citenamefont{Ashcroft and Mermin}(1976)}]{AM76}
\bibinfo{author}{\bibfnamefont{N.~W.} \bibnamefont{Ashcroft}} \bibnamefont{and}
  \bibinfo{author}{\bibfnamefont{N.~D.} \bibnamefont{Mermin}},
  \emph{\bibinfo{title}{Solid state physics}} (\bibinfo{publisher}{Saunders
  College Publ. Philadelphia}, \bibinfo{year}{1976}).

\bibitem[{\citenamefont{Wintersgill et~al.}(1979)\citenamefont{Wintersgill,
  Fontanella, Andeen, and Schuele}}]{WFA79}
\bibinfo{author}{\bibfnamefont{M.}~\bibnamefont{Wintersgill}},
  \bibinfo{author}{\bibfnamefont{J.}~\bibnamefont{Fontanella}},
  \bibinfo{author}{\bibfnamefont{C.}~\bibnamefont{Andeen}}, \bibnamefont{and}
  \bibinfo{author}{\bibfnamefont{D.}~\bibnamefont{Schuele}},
  \bibinfo{journal}{J. Appl. Phys.} \textbf{\bibinfo{volume}{50}},
  \bibinfo{pages}{8259} (\bibinfo{year}{1979}).

\bibitem[{\citenamefont{Butler et~al.}(2001)\citenamefont{Butler, Zhang,
  Schulthess, and MacLaren}}]{BZS01}
\bibinfo{author}{\bibfnamefont{W.~H.} \bibnamefont{Butler}},
  \bibinfo{author}{\bibfnamefont{X.-G.} \bibnamefont{Zhang}},
  \bibinfo{author}{\bibfnamefont{T.~C.} \bibnamefont{Schulthess}},
  \bibnamefont{and} \bibinfo{author}{\bibfnamefont{J.~M.}
  \bibnamefont{MacLaren}}, \bibinfo{journal}{Phys. Rev. B}
  \textbf{\bibinfo{volume}{63}}, \bibinfo{pages}{054416}
  (\bibinfo{year}{2001}).

\bibitem[{\citenamefont{Schubert et~al.}(2000)\citenamefont{Schubert, Tiwald,
  and Herzinger}}]{STH00}
\bibinfo{author}{\bibfnamefont{M.}~\bibnamefont{Schubert}},
  \bibinfo{author}{\bibfnamefont{T.~E.} \bibnamefont{Tiwald}},
  \bibnamefont{and} \bibinfo{author}{\bibfnamefont{C.~M.}
  \bibnamefont{Herzinger}}, \bibinfo{journal}{Phys. Rev. B}
  \textbf{\bibinfo{volume}{61}}, \bibinfo{pages}{8187} (\bibinfo{year}{2000}).

\bibitem[{\citenamefont{Bersch et~al.}(2008)\citenamefont{Bersch, Rangan,
  Bartynski, Garfunkel, and Vescovo}}]{BRB08}
\bibinfo{author}{\bibfnamefont{E.}~\bibnamefont{Bersch}},
  \bibinfo{author}{\bibfnamefont{S.}~\bibnamefont{Rangan}},
  \bibinfo{author}{\bibfnamefont{R.~A.} \bibnamefont{Bartynski}},
  \bibinfo{author}{\bibfnamefont{E.}~\bibnamefont{Garfunkel}},
  \bibnamefont{and} \bibinfo{author}{\bibfnamefont{E.}~\bibnamefont{Vescovo}},
  \bibinfo{journal}{Phys. Rev. B} \textbf{\bibinfo{volume}{78}},
  \bibinfo{pages}{085114} (\bibinfo{year}{2008}).

\bibitem[{\citenamefont{Perevalov et~al.}(2007)\citenamefont{Perevalov,
  Shaposhnikov, Gritsenko, Wong, Han, and Kim}}]{PSG07}
\bibinfo{author}{\bibfnamefont{T.~V.} \bibnamefont{Perevalov}},
  \bibinfo{author}{\bibfnamefont{A.~V.} \bibnamefont{Shaposhnikov}},
  \bibinfo{author}{\bibfnamefont{V.~A.} \bibnamefont{Gritsenko}},
  \bibinfo{author}{\bibfnamefont{H.}~\bibnamefont{Wong}},
  \bibinfo{author}{\bibfnamefont{J.~H.} \bibnamefont{Han}}, \bibnamefont{and}
  \bibinfo{author}{\bibfnamefont{C.~W.} \bibnamefont{Kim}},
  \bibinfo{journal}{JETP Letters} \textbf{\bibinfo{volume}{85}},
  \bibinfo{pages}{165} (\bibinfo{year}{2007}).

\bibitem[{\citenamefont{French}(1990)}]{French90}
\bibinfo{author}{\bibfnamefont{R.~H.} \bibnamefont{French}},
  \bibinfo{journal}{J. Am. Ceram. Soc.} \textbf{\bibinfo{volume}{73}},
  \bibinfo{pages}{477} (\bibinfo{year}{1990}).

\bibitem[{\citenamefont{Glinka and Jaroniec}(1997)}]{GJ97}
\bibinfo{author}{\bibfnamefont{Y.~D.} \bibnamefont{Glinka}} \bibnamefont{and}
  \bibinfo{author}{\bibfnamefont{M.}~\bibnamefont{Jaroniec}},
  \bibinfo{journal}{J. Appl. Phys.} \textbf{\bibinfo{volume}{82}},
  \bibinfo{pages}{3499} (\bibinfo{year}{1997}).

\bibitem[{\citenamefont{Ludeke et~al.}(1998)\citenamefont{Ludeke, Wen, and
  Schenk}}]{LWS98}
\bibinfo{author}{\bibfnamefont{R.}~\bibnamefont{Ludeke}},
  \bibinfo{author}{\bibfnamefont{H.~J.} \bibnamefont{Wen}}, \bibnamefont{and}
  \bibinfo{author}{\bibfnamefont{A.}~\bibnamefont{Schenk}},
  \bibinfo{journal}{Appl. Phys. Lett.} \textbf{\bibinfo{volume}{73}},
  \bibinfo{pages}{1221} (\bibinfo{year}{1998}).

\bibitem[{\citenamefont{Chanana}(2011)}]{Chanana11}
\bibinfo{author}{\bibfnamefont{R.~K.} \bibnamefont{Chanana}},
  \bibinfo{journal}{J. Appl. Phys.} \textbf{\bibinfo{volume}{109}},
  \bibinfo{pages}{104508} (\bibinfo{year}{2011}).

\bibitem[{\citenamefont{van Zeghbroeck}(2010)}]{Zeghbroeck10}
\bibinfo{editor}{\bibfnamefont{B.}~\bibnamefont{van Zeghbroeck}}, ed.,
  \emph{\bibinfo{title}{Principles of Semiconductor Devices and
  Heterojunctions}} (\bibinfo{publisher}{Prentice Hall}, \bibinfo{year}{2010}).

\end{thebibliography}
%\bibliographystyle{apsrev}

\end{document}